\documentclass[11pt]{article}

\usepackage{graphics,graphicx,color,float}
\usepackage{amsmath}
\usepackage{amssymb}
\usepackage{amsfonts}
\usepackage{amsthm, hyperref}
\usepackage[usenames,dvipsnames]{xcolor}
\usepackage{geometry}
\usepackage{tikz-cd}

\newtheorem{fact}{Fact}

\newcommand{\beq}{\begin{equation}}
\newcommand{\enq}{\end{equation}}
\newcommand{\bel}{\begin{lemma}}
\newcommand{\enl}{\end{lemma}}
\newcommand{\bet}{\begin{theorem}}
\newcommand{\ent}{\end{theorem}}

\newcommand{\E}{\mathbb{E}}
\newcommand{\Tr}{\mathrm{Tr}}
\newcommand{\ketbra}[1]{|#1\rangle\langle#1|}

\newcommand{\eps}{\varepsilon}

\newcommand{\id}{\ensuremath{\mathrm{I}}}
\newcommand{\rA}{\ensuremath{2^{R_A}}}
\newcommand{\rB}{\ensuremath{2^{R_B}}} 
\newcommand{\rApls}{\ensuremath{2^{R_A+r_A}}}
\newcommand{\rBpls}{\ensuremath{2^{R_B+r_B}}} 
\newcommand{\swapp}{\mathrm{SWAP}}

\newcommand*{\cA}{\mathcal{A}}

\newcommand*{\cH}{\mathcal{H}}

\newcommand*{\cB}{\mathcal{B}}
\newcommand*{\cD}{\mathcal{D}}

\newcommand*{\cE}{\mathcal{E}}

\newcommand{\cP}{\mathcal{P}}

\newcommand{\suppress}[1]{}

\newcommand{\defeq}{\ensuremath{ \stackrel{\mathrm{def}}{=} }}
\newcommand{\F}{\mathrm{F}}
\newcommand{\Pur}{\mathrm{P}}
\newcommand {\br} [1] {\ensuremath{ \left( #1 \right) }}

\newcommand {\minusspace} {\: \! \!}
\newcommand {\smallspace} {\: \!}
\newcommand {\fn} [2] {\ensuremath{ #1 \minusspace \br{ #2 } }}
\newcommand {\ball} [2] {\fn{\mathcal{B}^{#1}}{#2}}
\newcommand {\relent} [2] {\fn{\mathrm{D}}{#1 \middle\| #2}}
\newcommand {\dmax} [2] {\fn{\mathrm{D}_{\max}}{#1 \middle\| #2}}

\newcommand {\dmaxeps} [3] {\fn{\mathrm{D}^{#3}_{\max}}{#1 \middle\| #2}}
\newcommand {\dmineps} [3] {\fn{\mathrm{D}^{#3}_{\mathrm{H}}}{#1 \middle\| #2}}
\newcommand {\dseps} [3] {\fn{\mathrm{D}^{#3}_{s}}{#1 \middle\| #2}}
\newcommand {\dsepsalt} [3] {\fn{\tilde{\mathrm{D}}^{#3}_{s}}{#1 \middle\| #2}}
\newcommand {\mutinf} [2] {\fn{\mathrm{I}}{#1 \smallspace : \smallspace #2}}
\newcommand {\trimutinf} [3] {\fn{\mathrm{I}}{#1 \smallspace : \smallspace #2 \smallspace : \smallspace #3}}
\newcommand {\imax}{\ensuremath{\mathrm{I}_{\max}}}
\newcommand {\imaxeps} [1] {\ensuremath{\mathrm{\tilde{I}}^{#1}_{\max}}}
\newcommand {\imaxepss} [1] {\ensuremath{\mathrm{I}^{#1}_{\max}}}
\newcommand {\imaxepsbeta} [2] {\ensuremath{\mathrm{\bar{I}}^{#1,#2}_{\max}}}
\newcommand {\condmutinf} [3] {\mutinf{#1}{#2 \smallspace \middle\vert \smallspace #3}}

\newcommand {\rmS}{\mathrm{S}}

\newcommand{\bra}[1]{\langle #1|}
\newcommand{\ket}[1]{|#1 \rangle}

\mathchardef\mhyphen="2D

\newcommand{\tj}{\tilde{j_2}}
\newcommand{\tk}{\tilde{k_2}}

\makeatletter
\newcommand*{\rom}[1]{\expandafter\@slowromancap\romannumeral #1@}
\makeatother

\mathchardef\mhyphen="2D

\newtheorem{claim}{Claim}

\newtheorem{theorem}{Theorem}
\newtheorem{lemma}{Lemma}

\begin {document}
\title{A generalized quantum Slepian-Wolf} 
\author{
Anurag Anshu\footnote{Centre for Quantum Technologies, National University of Singapore, Singapore. \texttt{a0109169@u.nus.edu}} \qquad
Rahul Jain\footnote{Centre for Quantum Technologies, National University of Singapore and MajuLab, UMI 3654, 
Singapore. \texttt{rahul@comp.nus.edu.sg}} \qquad 
Naqueeb Ahmad Warsi\footnote{Centre for Quantum Technologies, National University of Singapore and School of Physical and Mathematical Sciences, Nanyang Technological University, Singapore and IIITD, Delhi. \texttt{warsi.naqueeb@gmail.com}} 
}

\date{}
\maketitle

%\date{}
\maketitle

\begin{abstract}
In this work we consider a quantum generalization of the task considered by Slepian and Wolf~\cite{SlepianW73} regarding distributed source compression. In our task Alice, Bob, Charlie and Reference share a joint pure state. Alice and Bob wish to send a part of their respective systems to Charlie without collaborating with each other. We give achievability bounds for this task in the one-shot setting and provide the asymptotic and i.i.d. analysis in the case when there is no side information with Charlie. 

Our result implies the result of  Abeyesinghe, Devetak, Hayden and Winter in~\cite{ADHW09} who studied  a special case of this problem. As another special case wherein Bob holds trivial registers, we recover the result of Devetak and Yard \cite{Devatakyard} regarding quantum state redistribution.
\end{abstract}

\section{Introduction}

In information theory, one of the most fundamental problems is the task of source-compression. The answer to this problem was given by Shannon in his celebrated work \cite{Shannon}. Slepian and Wolf, in their work \cite{SlepianW73}, studied this task in the distributed network setting, which consists of three parties Alice ($X_1,X_2\ldots X_n$ ), Bob ($Y_1,Y_2\ldots Y_n$) and Charlie, where $(X_1,Y_1), (X_2,Y_2), \ldots (X_n,Y_n)$ are pairs of independent and identically distributed correlated random variables.  The goal here is that Alice needs to communicate $(X_1,X_2,\ldots X_n)$ to Charlie and similarly, Bob needs to communicate $(Y_1,Y_2,\ldots Y_n)$ to Charlie. Furthermore, Alice and Bob do not collaborate. From Shannon's result, one can easily see that the amount of total communication needed to accomplish this task is $nH(X) + nH(Y)$. However, the surprising feature of the result of Slepian and Wolf is that the amount of total communication only needs to be $nH(XY)$. Furthermore, their result implies that there is a trade-off on the amount of communication between (Alice, Charlie) and (Bob, Charlie). 

The quantum version of this problem was studied by Abeyesinghe, Devetak, Hayden and Winter in~\cite{ADHW09}. In this setting, there are four parties, Alice (M), Bob (N), Charlie and Reference (R), where Reference serves as a purifying system for Alice and Bob. The goal is that Alice needs to communicate the register $M$ to Charlie and Bob needs to communicate the register $N$ to Charlie, such that the final quantum state between Reference and Charlie is close to the original pure state between Reference, Alice and Bob. The work \cite{ADHW09} studied above task in the asymptotic and i.i.d setting. The authors introduced a protocol termed \textit{Fully Quantum Slepian-Wolf} and combined it with Schumacher's compression \cite{Schumacher95} (using the notion of \textit{time-sharing}) to obtain a rate pair.

The emerging framework of one-shot information theory is providing a new perspective on data compression and channel coding and is relevant in the practical scenarios. It has also led to new insights into the conceptual details of information theoretic protocols. This is largely because the notational complications arising due to many copies of the state are no longer present (although we note that the asymptotic and i.i.d setting also has its own conveniences). One-shot information theory has also found applications in both classical communication complexity \cite{HJMR10,BravermanR11} and quantum communication complexity \cite{Dave14}. Many quantum tasks have been formulated in their one-shot setting, such as quantum state merging (\cite{Berta09, Renner11}, originally introduced in \cite{horodecki07}) and quantum state redistribution (\cite{Berta14,ADJ14,AnshuJW17SR}, originally introduced in \cite{Devatakyard, YardD09}). A one-shot version of the distributed source compression with multiple senders was studied in the work \cite{DutilH10}, where the authors considered the {\em entanglement consumption} (in place of the quantum communication cost) of the protocol. 

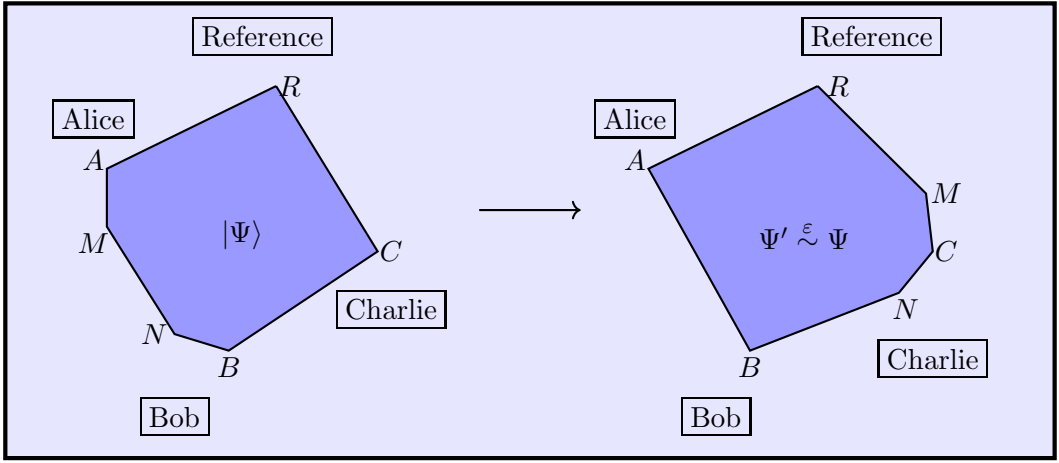
\begin{figure}[ht]
\centering
\begin{tikzpicture}[xscale=0.9,yscale=1.1]
%\draw[help lines] (0,0) grid (8,6);

\draw[ultra thick, fill = blue!10!white] (-4.5,6) rectangle (11,0.5);

\draw[thick, fill=blue!40!white] (-0.5,5) -- (-3,4) -- (-3,3.3) -- (-2,2)-- (-1.2,1.8)-- (1,3) -- (-0.5,5);
\node at (-0.3,5) {$R$};
\node at (-3.2,4.1) {$A$};
\node at (-3.2,3.1) {$M$};
\node at (-2.3,2) {$N$};
\node at (-1.2,1.6) {$B$};
\node at (1.2,3) {$C$};

\node at (-1,3.2) {$\ket{\Psi}$};

\node at (-0.7,5.6) {\fbox{Reference}};
\node at (-3.2,4.6) {\fbox{Alice}};
\node at (-2,1) {\fbox{Bob}};
\node at (1.2,2.3) {\fbox{Charlie}};

\draw[->, thick] (2.5,3.5) -- (4,3.5);

\draw[thick, fill=blue!40!white] (7.5,5) -- (5,4) -- (6.5,1.8)-- (8.7,2.5)-- (9.2,3)-- (9.1,3.7) -- (7.5,5);
\node at (7.8,5) {$R$};
\node at (4.8,4.1) {$A$};
\node at (9.4,3.7) {$M$};
\node at (8.8,2.3) {$N$};
\node at (6.5,1.6) {$B$};
\node at (9.4,3) {$C$};

\node at (7.3,3.2) {$\Psi' \overset{\eps}\sim \Psi$};

\node at (8.3,5.6) {\fbox{Reference}};
\node at (4.8,4.6) {\fbox{Alice}};
\node at (6,1) {\fbox{Bob}};
\node at (9.2,1.7) {\fbox{Charlie}};

\end{tikzpicture}
\caption{\small Alice (AM), Bob (BN), Charlie (C) and Reference (R) share a joint pure quantum state $\Psi$. The goal is that Alice needs to communicate the register $M$ to Charlie and Bob needs to communicate the register $N$ to Charlie, such that the final quantum state $\Psi'$ between Reference (R), Alice (A), Bob (B) and Charlie (CMN) is close to the original pure state between the parties. Shared entanglement is allowed only between (Alice, Charlie) and (Bob,Charlie) respectively.}
 \label{fig:qslepianwolf}
\end{figure}

Given the importance of one-shot information theory, in this work we consider the one-shot version of the problem studied in \cite{ADHW09}. To capture a more general scenario, along with the registers $M,N$ we also allow Alice, Bob and Charlie to have additional registers $A,B,C$ respectively. Thus, our setting is as follows, depicted in Figure \ref{fig:qslepianwolf}. 

\vspace{0.1in}

\noindent {\bf Task 1:} Alice (AM), Bob (BN), Charlie (C) and Reference (R) share a joint pure quantum state. The goal is that Alice needs to communicate the register $M$ to Charlie and Bob needs to communicate the register $N$ to Charlie, such that the final quantum state between Reference (R), Alice (A), Bob (B) and Charlie (CMN) is close to the original pure state between the parties. We allow pre-shared entanglement between (Alice, Charlie) and (Bob, Charlie) respectively.

This task is a natural generalization of the aforementioned task considered in \cite{ADHW09} and also extends the well studied problem of quantum state redistribution \cite{Devatakyard,YardD09}.  A special case when $A$ is trivial was considered by \cite{HsiehW2015} in which they studied the trade-off between the amount of entanglement consumed between Alice and Charlie and the communication between Bob and Charlie. The task is also natural for the well studied simultaneous message passing model \cite{GavinskyRW08, JainKlauck09, Gavinsky17} in quantum communication complexity. Furthermore, in the setting of quantum communication complexity with three parties, all the parties receive an input and hence can have some side information about the messages from other parties. This is partially captured by above task, a caveat being that the restriction on shared entanglement may not be necessary in general in quantum communication complexity. A special case of the above task, the quantum state redistribution, has found important recent applications in quantum communication complexity \cite{Dave14}.

\vspace{0.1in}

In our results, we shall also consider a \textit{time-reversed} version of the above task, as stated below. 

\vspace{0.1in}

\noindent {\bf Task 2:} Alice (A), Bob (B), Charlie (CMN) and Reference (R) share a joint pure quantum state. The goal is that Charlie needs to communicate the register $M$ to Alice and the register $N$ to Bob, such that the final quantum state between Reference (R), Alice (M), Bob (N) and Charlie (C) is close to the original pure state between the parties. We allow pre-shared entanglement between (Alice, Charlie) and (Bob, Charlie) respectively.

The motivation to study this task comes from the fact that near-optimal one-shot bounds on the entanglement assisted quantum communication cost of quantum state merging have been obtained by constructing protocols for its time reversed version, quantum state splitting \cite{Renner11, ADJ14}. 
 
\vspace{0.1in}

\noindent {\bf Our Results:} Our one-shot result is stated as Theorem~\ref{thm:task1}. We emphasize upon two main ingredients: 
\begin{itemize}
\item First is that the achievable rate region appears in terms of the max-relative entropy and the hypothesis testing relative entropy. 
\item Second is that the achievable rate region is a union of a family of achievable rate regions, each characterized by a quantum state that is close to original state $\Psi$ and satisfies some max-relative entropy constraints. 
\end{itemize}

Using this, we are able to obtain the following achievable rate region for Task $1$ in the asymptotic and i.i.d setting, when the register $C$ is trivial: 
\begin{eqnarray*}
R_{A\to C} &\geq& \frac{1}{2}\left(\mutinf{RAB}{M} - \mutinf{A}{M} \right),\\
R_{B\to C} &\geq& \frac{1}{2}\left(\mutinf{RAB}{N} - \mutinf{B}{N} \right), \\
R_{A\to C}+R_{B\to C}&\geq& \frac{1}{2}\bigg(\trimutinf{RAB}{M}{N}-\mutinf{A}{M} -\mutinf{B}{N}\bigg),
\end{eqnarray*} 
where $R_{A\to C}$ is the rate of quantum communication from Alice to Charlie, $R_{B\to C}$ is the rate of quantum communication from Bob to Charlie and all the information theoretic quantities calculated above are with respect to the state $\Psi_{RAMBN}$ shared between Alice, Bob and Reference. Note that we have used a tripartite version of the mutual information, formally defined in Section \ref{sec:prelim}. 

An immediate consequence of the above result is the rate pair obtained for the task considered in \cite{ADHW09}, with the registers $A,B$ being trivial. Moreover, if the registers $B,N$ are trivial in the original task, then the task reduces to that of quantum state redistribution. In this case, the result of Theorem~\ref{thm:task1} also reproduces the bound given in \cite{Devatakyard,YardD09} for quantum state redistribution in the asymptotic and i.i.d. setting.

\vspace{0.1in}

\noindent {\bf Converse bounds:} There are two challenges for obtaining the one-shot converse rate region for our tasks. First is that a matching converse bound for the task of quantum state redistribution, which is a special case of our tasks, is not known in the one-shot setting and is a major open question in quantum information theory. Second, a matching converse for the task considered in \cite{ADHW09} is not known even in the asymptotic and i.i.d. setting (as discussed in \cite[Section $10$]{ADHW09}). We are not able to solve any of these challenges,  but are able to show a matching one-shot converse for the achievable rate region of Task $2$, in the special case where registers $A,B$ are trivial (more details appear in Section \ref{sec:converse}). One might be tempted to suggest that this should imply a matching converse for Task $1$ in the special case where registers $A,B$ are trivial (in analogy with quantum state merging and quantum state splitting). Unfortunately this is not the case, since a general protocol for Task $1$ might start with Alice and Bob distilling out a pure state on their registers, and then proceeding with a potentially easier communication task. This problem was already recognized in \cite[Section $10$]{ADHW09}, which led to a gap between their achievable rate region and their converse. We point out that this problem does not arise in Task $2$, as Alice and Bob are not allowed to share entanglement before the protocol starts. 

\vspace{0.1in}

\noindent {\bf Techniques:} Along with the inherent challenges of one-shot information theory, an additional challenge for extending the result of \cite{ADHW09} is the absence of the notion of time sharing in the one-shot case. The idea of time-sharing is as follows: given two rates $R=(R_1,R_2)$ and $R'= (R'_1,R'_2)$ at which Alice and Bob can communicate to Charlie, one can construct a protocol which achieves the rate $\alpha R + (1-\alpha)R'$ by using the first protocol for the first $\alpha n$ copies and using the second protocol for the last $(1 - \alpha)n$ copies (see \cite[Page 534]{CoverT91}). 

It is clear that this technique cannot extend to the one-shot setting which considers just one copy of the input state. We overcome the obstacle of time sharing in the one-shot case by using the technique of \textit{convex-split}~\cite{ADJ14} along with \textit{position-based decoding}~\cite{AnshuJW17CC}. The convex-split technique allows one party to prepare a convex combination of states on the registers of other party, if the first party holds a purification of the registers of the second party.  The concept of position-based decoding is essentially that of hypothesis testing on a global state. 

The technical contribution of this work resides in two aspects. First is that we prove a new version of convex-split lemma~\cite[Page 3]{ADJ14}, which we refer to as \textit{tripartite} convex-split lemma, which requires Charlie to prepare a convex combination of quantum states shared between three parties Reference, Alice and Bob. We prove the sufficient conditions which allow Charlie to prepare such convex combination with small error. The second technical contribution is in our asymptotic and i.i.d. analysis of the one-shot bounds. It can be seen that the time-sharing technique, along with the quantum state redistribution protocol of \cite{Devatakyard,YardD09}, obtains the asymptotic and i.i.d. achievability result mentioned above \footnote{The extremal points of the achievable rate region are $(R_{A\to C}, R_{B\to C}) = (\frac{1}{2}\condmutinf{RB}{M}{NC}, \frac{1}{2}\condmutinf{RAM}{N}{C})$ and $(R_{A\to C}, R_{B\to C}) = (\frac{1}{2}\condmutinf{RBN}{M}{C}, \frac{1}{2}\condmutinf{RA}{N}{MC})$. The first can be achieved by Bob sending $N$ to Charlie using quantum state redistribution, followed by Alice sending $M$ to Charlie, again using quantum state redistribution. Second can be achieved in analogous fashion. Any rate pair can then be achieved by time sharing between these two protocols.}. Since our one-shot result has no time-sharing involved, we provide an explicit analysis of our bound when there are many independent copies of the state $\Psi$ shared between the parties, in the case where register $C$ is absent. For this, we exploit several properties of the quantum information spectrum relative entropy (introduced in \cite{HyashiN03, NagaokaH07}; the classical information spectrum approach originated in \cite{HanV93}) to show the existence of a quantum state that is close to the original state $\Psi$ and satisfies several max-entropy constraints on the reduced systems (given explicitly in the statement of Theorem \ref{thm:asymptotic}). A special case of this analysis has also appeared in the context of quantum channel coding for the quantum broadcast channel in \cite{AnshuJW17CC}, suggesting a wide applicability of the techniques developed in the proof of Theorem \ref{thm:asymptotic}.

\subsection*{Organization}
We provide our notations and useful facts in Section \ref{sec:prelim}. We discuss our achievability protocol in Section \ref{sec:main} and the asymptotic and i.i.d. bounds in Section \ref{sec:asymptoticiid}. We discuss a converse result in Section \ref{sec:converse}. We prove the tripartite version of convex-split lemma in Appendix \ref{sec:convexsplit} and give details of the asymptotic and i.i.d. analysis in Appendix \ref{sec:asymptote}.

\section{Quantum information theory}
\label{sec:prelim}
Consider a finite dimensional Hilbert space $\cH$ endowed with an inner product $\langle \cdot, \cdot \rangle$ (in this paper, we only consider finite dimensional Hilbert-spaces). The $\ell_1$ norm of an operator $X$ on $\cH$ is $\| X\|_1\defeq\Tr\sqrt{X^{\dagger}X}$ and $\ell_2$ norm is $\| X\|_2\defeq\sqrt{\Tr XX^{\dagger}}$. A quantum state (or a density matrix or a state) is a positive semi-definite matrix on $\cH$ with trace equal to $1$. It is called {\em pure} if and only if its rank is $1$. A sub-normalized state is a positive semi-definite matrix on $\cH$ with trace less than or equal to $1$. Let $\ket{\psi}$ be a unit vector on $\cH$, that is $\langle \psi,\psi \rangle=1$.  With some abuse of notation, we use $\psi$ to represent the state and also the density matrix $\ketbra{\psi}$, associated with $\ket{\psi}$. Given a quantum state $\rho$ on $\cH$, {\em support of $\rho$}, called $\text{supp}(\rho)$ is the subspace of $\cH$ spanned by all eigenvectors of $\rho$ with non-zero eigenvalues.
 
A {\em quantum register} $A$ is associated with some Hilbert space $\cH_A$. Define $|A| \defeq \dim(\cH_A)$. Let $\mathcal{L}(A)$ represent the set of all linear operators on $\cH_A$. Let $\mathcal{P}(A)$ represent the set of all positive semidefinite operators on $\cH_A$. We denote by $\mathcal{D}(A)$, the set of quantum states on the Hilbert space $\cH_A$. State $\rho$ with subscript $A$ indicates $\rho_A \in \mathcal{D}(A)$. If two registers $A,B$ are associated with the same Hilbert space, we shall represent the relation by $A\equiv B$.  Composition of two registers $A$ and $B$, denoted $AB$, is associated with Hilbert space $\cH_A \otimes \cH_B$.  For two quantum states $\rho\in \mathcal{D}(A)$ and $\sigma\in \mathcal{D}(B)$, $\rho\otimes\sigma \in \mathcal{D}(AB)$ represents the tensor product (Kronecker product) of $\rho$ and $\sigma$. The identity operator on $\cH_A$ (and associated register $A$) is denoted $\id_A$. For any operator $O$ on $\cH_A$, we denote by $\{O\}_+$ the subspace spanned by non-negative eigenvalues of $O$ and by $\{O\}_-$ the subspace spanned by negative eigenvalues of $O$.  For a positive semidefinite operator $M\in \mathcal{P}(A)$, the largest and smallest non-zero eigenvalues of $M$ are denoted by $\lambda_{max}(M)$ and $\lambda_{min}(M)$, respectively.

Let $\rho_{AB} \in \mathcal{D}(AB)$. We define
$$ \rho_{B} \defeq \Tr_{A}\rho_{AB} \defeq \sum_i (\bra{i} \otimes \id_{B})\rho_{AB} (\ket{i} \otimes \id_{B}) ,$$
where $\{\ket{i}\}_i$ is an orthonormal basis for the Hilbert space $\cH_A$.
The state $\rho_B\in \mathcal{D}(B)$ is referred to as the marginal state of $\rho_{AB}$. Unless otherwise stated, a register missing from the subscript of a state will represent the partial trace over that register. Given a $\rho_A\in\mathcal{D}(A)$, a {\em purification} of $\rho_A$ is a pure state $\rho_{AB}\in \mathcal{D}(AB)$ such that $\Tr{B}{\rho_{AB}}=\rho_A$. A purification of a quantum state is not unique.

A quantum {map} $\cE: \mathcal{L}(A)\rightarrow \mathcal{L}(B)$ is a completely positive and trace preserving (CPTP) linear map (mapping states in $\mathcal{D}(A)$ to states in $\mathcal{D}(B)$). A {\em unitary} operator $U_A:\cH_A \rightarrow \cH_A$ is such that $U_A^{\dagger}U_A = U_A U_A^{\dagger} = \id_A$. An {\em isometry}  $V:\cH_A \rightarrow \cH_B$ is such that $V^{\dagger}V = \id_A$ and $VV^{\dagger} = \Pi_B$, for a projection $\Pi_B$ on $\cH_B$. The set of all unitary operations on register $A$ is denoted by $\mathcal{U}(A)$. For registers $A$ and $B$ with $|A|=|B|$, the operation that swaps these registers is $\swapp_{A,B}\defeq \sum_{i,j} \ket{i,j}\bra{j,i}$, for an arbitrary basis $\{\ket{i}\}_{i=1}^{|A|}, \{\ket{j}\}_{j=1}^{|B|}$ on $\cH_A, \cH_B$ respectively.

We shall consider the following information theoretic quantities. Let $\varepsilon \in (0,1)$. 
\begin{enumerate}
\item {\bf Fidelity} (\cite{Josza94}, see also \cite{uhlmann76}) For $\rho_A,\sigma_A \in \mathcal{D}(A)$, $$\F(\rho_A,\sigma_A)\defeq\|\sqrt{\rho_A}\sqrt{\sigma_A}\|_1.$$ For classical probability distributions $P = \{p_i\}$, $Q =\{q_i\}$, $$\F(P,Q)\defeq \sum_i \sqrt{p_i \cdot q_i}.$$
\item {\bf Purified distance} (\cite{GilchristLN05}) For $\rho_A,\sigma_A \in \mathcal{D}(A)$, $$\Pur(\rho_A,\sigma_A) = \sqrt{1-\F^2(\rho_A,\sigma_A)}.$$

\item {\bf $\varepsilon$-ball} For $\rho_A\in \mathcal{D}(A)$, $$\ball{\eps}{\rho_A} \defeq \{\rho'_A\in \mathcal{D}(A)|~\Pur(\rho_A,\rho'_A) \leq \varepsilon\}. $$ 
\item {\bf Von-Neumann entropy} (\cite{Neumann32}) For $\rho_A\in\mathcal{D}(A)$, $$\rmS(\rho_A) \defeq - \Tr(\rho_A\log\rho_A) .$$ 
\item {\bf Relative entropy} (\cite{umegaki1954}) For $\rho_A\in \mathcal{D}(A)$, $\sigma_A\in \mathcal{P}(A)$ such that $\text{supp}(\rho_A) \subset \text{supp}(\sigma_A)$, $$\relent{\rho_A}{\sigma_A} \defeq \Tr(\rho_A\log\rho_A) - \Tr(\rho_A\log\sigma_A) .$$ 

\item {\bf Max-relative entropy} (\cite{Datta09}) For $\rho_A,\sigma_A\in \mathcal{P}(A)$ such that $\text{supp}(\rho_A) \subset \text{supp}(\sigma_A)$, $$ \dmax{\rho_A}{\sigma_A}  \defeq  \inf \{ \lambda \in \mathbb{R} :  \rho_A \preceq 2^{\lambda} \sigma_A\}  .$$  
\item {\bf Smooth max-relative entropy} (\cite{Datta09}, see also \cite{Jain:2009}) For $\rho_A\in \mathcal{D}(A) ,\sigma_A\in \mathcal{P}(A)$ such that $\text{supp}(\rho_A) \subset \text{supp}(\sigma_A)$, $$ \dmaxeps{\rho_A}{\sigma_A}{\eps}  \defeq  \sup_{\rho'_A\in \ball{\eps}{\rho_A}} \dmax{\rho_A'}{\sigma_A}  .$$  
\item {\bf Hypothesis testing relative entropy} (\cite{BuscemiD10}, see also \cite{HyashiN03}) For $\rho_A\in \mathcal{D}(A) ,\sigma_A\in \mathcal{P}(A)$, $$ \dmineps{\rho_A}{\sigma_A}{\eps}  \defeq  \sup_{0\preceq \Pi\preceq \id, \Tr(\Pi\rho_A)\geq 1-\eps}\log\left(\frac{1}{\Tr(\Pi\sigma_A)}\right).$$  

\item {\bf Information spectrum relative entropy} (\cite{HyashiN03, NagaokaH07}) For $\rho_A\in \mathcal{D}(A) ,\sigma_A\in \mathcal{P}(A)$ such that $\text{supp}(\rho_A) \subset \text{supp}(\sigma_A)$, $$ \dseps{\rho_A}{\sigma_A}{\eps}  \defeq  \sup \{R: \Tr(\rho_A\{\rho_A-2^R\sigma_A\}_{+}) \geq 1-\eps \}  .$$  

\item {\bf Information spectrum relative entropy [Alternate definition]} For $\rho_A\in \mathcal{D}(A) ,\sigma_A\in \mathcal{P}(A)$ such that $\text{supp}(\rho_A) \subset \text{supp}(\sigma_A)$, $$ \dsepsalt{\rho_A}{\sigma_A}{\eps}  \defeq  \inf \{R: \Tr(\rho_A\{\rho_A-2^R\sigma_A\}_{-}) \geq 1-\eps \}  .$$  

\item {\bf Mutual information} For $\rho_{AB}\in \mathcal{D}(AB)$, $$\mutinf{A}{B}_{\rho}\defeq \rmS(\rho_A) + \rmS(\rho_B)-\rmS(\rho_{AB}) = \relent{\rho_{AB}}{\rho_A\otimes\rho_B}.$$

\item {\bf tripartite mutual information} For $\rho_{ABC}\in \mathcal{D}(ABC)$, $$\trimutinf{A}{B}{C}_{\rho}\defeq \rmS(\rho_A) + \rmS(\rho_B)+ \rmS(\rho_C)-\rmS(\rho_{ABC})$$ $$= \relent{\rho_{ABC}}{\rho_A\otimes\rho_B\otimes \rho_C}.$$

\suppress{
\item {\bf Max-information} For $\rho_{AB}\in \mathcal{D}(AB)$, define 
$$\imax(A:B)_{\rho} = \dmax{\rho_{AB}}{\rho_A\otimes \rho_B}.$$

\item {\bf Smooth max-information} For $\rho_{AB}\in \mathcal{D}(AB)$, define 
$$\imaxepss{\eps}(A:B)_{\rho} = \min_{\rho'\in\ball{\eps}{\rho}}\imax(A:B)_{\rho'} .$$

\item {\bf Smooth max-information [Alternate definition]} For $\rho_{AB}\in \mathcal{D}(AB)$, define 
$$\imaxeps{\eps}(A:B)_{\rho} = \min_{\rho'\in\ball{\eps}{\rho}}\dmax{\rho'_{AB}}{\rho'_A\otimes \rho_B}.$$

\item {\bf Restricted smooth max-information} For $\rho_{AB}\in \mathcal{D}(AB)$, define 
$$\imaxepsbeta{\eps}{\delta}(A:B)_{\rho} = \min_{\rho'\in\ball{\eps}{\rho}: \rho'_A\leq (1+\delta)\rho_A, \rho'_B\leq (1+\delta)\rho_B}\dmax{\rho'_{AB}}{\rho_A\otimes \rho_B}.$$

\item {\bf Conditional mutual information} For $\rho_{ABC}\in\mathcal{D}(ABC)$, $$\condmutinf{A}{B}{C}_{\rho}\defeq \mutinf{A}{BC}_{\rho}-\mutinf{A}{C}_{\rho}.$$
\item {\bf Max-information}  For $\rho_{AB}\in \mathcal{D}(AB)$, $$ \imax(A:B)_{\rho} \defeq   \inf_{\sigma_{B}\in \mathcal{D}(B)}\dmax{\rho_{AB}}{\rho_{A}\otimes\sigma_{B}} .$$
\item {\bf Smooth max-information} For $\rho_{AB}\in \mathcal{D}(AB)$,  $$\imaxeps(A:B)_{\rho} \defeq \inf_{\rho'\in \ball{\eps}{\rho}} \imax(A:B)_{\rho'} .$$	
}
\end{enumerate}

We will use the following facts. 
\begin{fact}[Triangle inequality for purified distance,~\cite{Tomamichel12, GilchristLN05}]
\label{fact:trianglepurified}
For quantum states $\rho_A, \sigma_A, \tau_A\in \mathcal{D}(A)$,
$$\Pur(\rho_A,\sigma_A) \leq \Pur(\rho_A,\tau_A)  + \Pur(\tau_A,\sigma_A) . $$ 
\end{fact}
\suppress{
\begin{fact}[\cite{stinespring55}](\textbf{Stinespring representation})\label{stinespring}
Let $\E(\cdot): \mathcal{L}(A)\rightarrow \mathcal{L}(B)$ be a quantum operation. There exists a register $C$ and an unitary $U\in \mathcal{U}(ABC)$ such that $\E(\omega)=\Tr_{A,C}\br{U (\omega  \otimes \ketbra{0}^{B,C}) U^{\dagger}}$. Stinespring representation for a channel is not unique. 
\end{fact}
}
\begin{fact}[Monotonicity under quantum operations, \cite{barnum96},\cite{lindblad75}]
	\label{fact:monotonequantumoperation}
For quantum states $\rho$, $\sigma \in \mathcal{D}(A)$, and quantum operation $\cE(\cdot):\mathcal{L}(A)\rightarrow \mathcal{L}(B)$, it holds that
\begin{align*}
	&\dmax{\rho}{\sigma} \geq \dmax{\cE(\rho)}{\cE(\sigma)},\\  &\F(\cE(\rho),\cE(\sigma)) \geq \F(\rho,\sigma),  \\ &\dmineps{\rho}{\sigma}{\eps} \geq \dmineps{\cE(\rho)}{\cE(\sigma)}{\eps}.
\end{align*}
In particular, for bipartite states $\rho_{AB},\sigma_{AB}\in \mathcal{D}(AB)$, it holds that
\begin{align*}
	&\dmax{\rho_{AB}}{\sigma_{AB}} \geq \dmax{\rho_A}{\sigma_A}, \\ &\F(\rho_{A},\sigma_{A}) \geq \F(\rho_{AB},\sigma_{AB}), \\ &\dmineps{\rho_{AB}}{\sigma_{AB}}{\eps} \geq \dmineps{\rho_A}{\sigma_A}{\eps}.
\end{align*}
\end{fact}

\begin{fact}[Uhlmann's Theorem, \cite{uhlmann76}]
\label{uhlmann}
Let $\rho_A,\sigma_A\in \mathcal{D}(A)$. Let $\rho_{AB}\in \mathcal{D}(AB)$ be a purification of $\rho_A$ and $\ket{\sigma}_{AC}\in\mathcal{D}(AC)$ be a purification of $\sigma_A$. There exists an isometry $V: C \rightarrow B$ such that,
 $$\F(\ketbra{\theta}_{AB}, \ketbra{\rho}_{AB}) = \F(\rho_A,\sigma_A) ,$$
 where $\ket{\theta}_{AB} = (\id_A \otimes V) \ket{\sigma}_{AC}$.
\end{fact}

Following fact implies the Pinsker's inequality.

\begin{fact}[Lemma 5,\cite{Jain:2003a}]
\label{pinsker}
For quantum states $\rho_A,\sigma_A\in\mathcal{D}(A)$, 
$$\F(\rho,\sigma) \geq 2^{-\frac{1}{2}\relent{\rho}{\sigma}}.$$
\end{fact}

\begin{fact}[Lemma B.7, \cite{Renner11}]
\label{fact:dmaxup}
For a quantum state $\rho_{AB} \in \mathcal{D}(AB)$, it holds that $\dmax{\rho_{AB}}{\rho_A\otimes \frac{\id_B}{|B|}} \leq 2\log|B|$.
\end{fact}

\begin{fact}[Gentle measurement lemma,\cite{Winter:1999,Ogawa:2002}]
\label{gentlelemma}
Let $\rho$ be a quantum state and $0\preceq A\preceq \id$ be an operator. Then 
$$\F(\rho, \frac{A\rho A}{\Tr(A^2\rho)})\geq \sqrt{\Tr(A^2\rho)}.$$
\end{fact}

\begin{lemma}
\label{gentlepovm}
Consider a pure quantum state $\ket{\rho}_{ORA} = \sum_i \sqrt{p_i}\ket{i}_O \ket{\rho^i}_{RA}$ and an isometry $\cA  = \sum_i P_i \otimes \ket{i}_{O'}$, such that $0 < P_i < \id_A, \sum_i P_i^2 = \id_A$. Define the state $\ket{\rho'}_{ORAO'}\defeq \sum_i \sqrt{p_i}\ket{i}_O\ket{\rho^i}_{RA}\ket{i}_{O'}$ and let $q_i \defeq \Tr(P^2_i \rho^i_A)$. Then it holds that
$$\F(\rho'_{ORAO'},\cA\rho_{ORA}\cA^{\dagger})\geq \sum_i p_iq_i.$$
\end{lemma}
\begin{proof}
Consider the state
$$\cA\ket{\rho}_{ORA} = \sum_{i,j}\sqrt{p_i}\ket{i}_O (\id_R\otimes P_j)\ket{\rho^i}_{RA}\ket{j}_{O'}.$$
We compute 
\begin{eqnarray*}
\F(\rho'_{ORAO'},\cA\rho_{ORA}\cA^{\dagger}) = &&  |(\sum_i \sqrt{p_i}\bra{i}_O\bra{\rho^i}_{RA}\bra{i}_{O'})\\ &&(  \sum_{i,j}\sqrt{p_i}\ket{i}_O (\id_R\otimes P_j)\ket{\rho^i}_{RA}\ket{j}_{O'})|\\ &&= | \sum_i p_i\bra{\rho^i}_{RA}(\id_R\otimes P_i)\ket{\rho^i}_{RA}|\\ &&= \sum_i p_i \Tr(P_i\rho^i_A) \geq \sum_i p_i \Tr(P^2_i\rho^i_A),
\end{eqnarray*}
where the last inequality follows from the fact that $P_i^2 \preceq P_i$, which is implied by $P_i \preceq \id_A$. This completes the proof by the definition of $q_i$.
\end{proof}

\begin{fact}[\cite{AnshuJW17SR}]
\label{closestatesmeasurement}
Let $\eps, \delta \in (0,1)$ such that $2\eps+\delta <1$. Let $\rho,\sigma$ be quantum states such that $\Pur(\rho,\sigma)\leq \eps$. Let $0\preceq \Pi\preceq \id$ be an operator such that $\Tr(\Pi\rho)\geq 1-\delta^2$. Then $\Tr(\Pi\sigma)\geq 1- (2\eps+\delta)^2$. If $\delta=0$, then $\Tr(\Pi\sigma) \geq 1-\eps^2$.
\end{fact}

\begin{fact}[Hayashi-Nagaoka inequality, \cite{HyashiN03}]
\label{haynag}
Let $0 \preceq S \preceq \id,T$ be positive semi-definite operators. Then 
$$\id - (S+T)^{-\frac{1}{2}}S(S+T)^{-\frac{1}{2}}\preceq 2(\id-S) + 4T.$$

\end{fact}

\section{Achievable rate region for distributed quantum source compression with side information}
\label{sec:main}

We define our tasks formally below.

\vspace{0.1in}

\noindent {\bf Task $1$:} There are four parties Alice, Bob, Charlie and Reference. Furthermore, Alice $(AM)$, Bob $(BN)$, Reference $(R)$ and Charlie $(C)$ share the joint pure state $\ket{\Psi}_{RAMBNC}$.  Alice and Bob wish to communicate their registers $M$ and $N$ to Charlie such that the final state shared between Alice $(A)$, Bob $(B)$, Reference $(R)$ and Charlie $(CMN)$ is $\Phi_{RABCMN}$ with the property that $\Pur(\Phi,\Psi)\leq \eps$, where $\eps \in (0,1)$ is an error parameter. To accomplish this task, Alice and Charlie are also allowed pre-shared entanglement. Similarly, Bob and Charlie are allowed the same. See Figure \ref{fig:qslepianwolf}.

\vspace{0.1in}

To accomplish Task 1, we will first consider the \textit{time-reversed} version defined as follows. 

\vspace{0.1in}

\noindent {\bf Task $2$:} There are four parties Alice, Bob, Charlie and Reference. Furthermore, Alice $(A)$, Bob $(B)$, Reference $(R)$ and Charlie $(CMN)$ share the joint pure state $\ket{\Psi}_{RABCMN}$.  Charlie wishes to communicate her register $M$ to Alice and $N$ to Bob such that the final state shared between Alice $(AM)$, Bob $(BN)$, Reference $(R)$ and Charlie $(C)$ is $\Phi_{RAMBNC}$ with the property that $\Pur(\Phi,\Psi)\leq \eps$, where $\eps \in (0,1)$ is an error parameter. To accomplish this task, Alice and Charlie are also allowed pre-shared entanglement. Similarly, Bob and Charlie are allowed the same. 

\subsection*{Main result: Achievable rate region for Task $2$}

\begin{theorem}
\label{thm:task2}
Fix $\eps_1,\eps_2,\delta\in (0,1)$ such that $\eps_1+5\eps_2+2\sqrt{\delta} <1$. Let Alice $(A)$, Bob $(B)$, Reference $(R)$ and Charlie $(CMN)$ share the pure state $\ket{\Psi}_{RABCMN}$. There exists an entanglement assisted quantum protocol with pre-shared entanglement of the form $\ket{\theta_1}\otimes\ket{\theta_2}$ (where $\ket{\theta_1}$ is shared between Alice, Charlie in some registers $E_{AC}$ and $\ket{\theta_2}$ is shared between Bob, Charlie in some registers $E_{BC}$), such that at the end of the protocol following properties hold.
\begin{itemize}
\item The global shared state is $\ket{\Phi}_{RAMBNCE'_{AC}E'_{BC}}$ with $R$ belonging to Reference, $(AM)$ belonging to Alice, $(BN)$ belonging to Bob, $C$ belonging to Charlie, $E'_{AC}$ belonging to (Alice, Charlie) and $E'_{BC}$ belonging to (Bob, Charlie).
\item There exist states $\ket{\theta'_1}_{E'_{AC}}$ and $\ket{\theta'_2}_{E'_{BC}}$ such that $\Pur(\ketbra{\Phi},\ketbra{\Psi}\otimes\ketbra{\theta'_1}\otimes\ketbra{\theta'_2})\leq \eps_1+5\eps_2+2\sqrt{\delta}$. 
\end{itemize}
The number of qubits that Charlie sends to Alice and Bob are $R_{C\to A}$ and $R_{C\to B}$ be respectively, where the pair $(R_{C\to A},R_{C\to B})$ lie in the union of the following achievable rate region: for every $\Psi'_{RABCMN} \in \ball{\eps_1}{\Psi_{RABCMN}}$ such that $\Psi'_{RAB} \preceq 2^{\delta}\Psi_{RAB}$ and states $\sigma_M,\omega_N$:

\begin{eqnarray*}
R_{C\to A} &\geq& \frac{1}{2}\bigg(\dmax{\Psi'_{RABM}}{\Psi_{RAB}\otimes \sigma_M} - \dmineps{\Psi_{AM}}{\Psi_A\otimes \sigma_M}{\eps_2^2} + \log\frac{1}{\eps_2^2\delta}\bigg),\\
R_{C\to B} &\geq& \frac{1}{2}\bigg(\dmax{\Psi'_{RABN}}{\Psi_{RAB}\otimes \omega_N} - \dmineps{\Psi_{BN}}{\Psi_B\otimes \omega_N}{\eps_2^2} + \log\frac{1}{\eps_2^2\delta}\bigg), \\
R_{C\to A}+R_{C\to B}&\geq& \frac{1}{2}\bigg(\dmax{\Psi'_{RABMN}}{\Psi_{RAB}\otimes\sigma_M\otimes\omega_N} -\dmineps{\Psi_{AM}}{\Psi_A\otimes \sigma_M}{\eps_2^2}\\ &&-\dmineps{\Psi_{BN}}{\Psi_B\otimes \omega_N}{\eps_2^2} +\log\frac{1}{\eps_2^2\delta}\bigg).\\
\end{eqnarray*}  
\end{theorem}

\begin{proof}
We divide our proof into the following steps. 
\vspace{0.1in}

\noindent {\bf 1. Quantum states and registers involved in the proof:} Fix $\Psi'_{RAB}\in \ball{\eps_1}{\Psi_{RAB}}$, states $\sigma_M,\omega_N$ and the pair $(R_{C\to A}, R_{C\to B})$ as mentioned in the lemma.  Let $R_A\defeq 2\cdot R_{C\to A}$ and $R_B\defeq 2\cdot R_{C\to B}$. Let $r_A$, $r_B$ be such that 
$$r_A \leq  \dmineps{\Psi_{AM}}{\Psi_A\otimes \sigma_M}{\eps_2^2} + 2\log\eps_2, $$ $$  r_B \leq  \dmineps{\Psi_{BN}}{\Psi_B\otimes \omega_N}{\eps_2^2} + 2\log\eps_2.$$ Let $\Pi^A_{AM}$ and $\Pi^B_{BN}$ be projectors achieving the optimum in the definitions of $\dmineps{\Psi_{AM}}{\Psi_A\otimes \sigma_M}{\eps_2^2}$ and $\dmineps{\Psi_{BN}}{\Psi_B\otimes \omega_N}{\eps_2^2}$ respectively. 

Introduce registers $M_1,M_2,\ldots M_{\rApls}$ such that for all $i$, $M_i \equiv M$ and $N_1,N_2,\ldots N_{\rBpls}$ such that for all $i$, $N_i \equiv N$. For brevity, we define $\sigma^{(-j)}\defeq\sigma_{M_1}\otimes\ldots\otimes\sigma_{M_{j-1}}\otimes\sigma_{M_{j+1}}\otimes\ldots \otimes\sigma_{M_{\rApls}}$ and $\omega^{(-k)}\defeq \omega_{N_1}\otimes\ldots\otimes\omega_{N_{k-1}}\otimes\omega_{N_{k+1}}\otimes\ldots \otimes\omega_{N_{\rBpls}}$. Consider the states, 

\begin{eqnarray*}
&&\mu_{RAB M_1 \ldots M_{\rApls} N_1 \ldots N_{\rBpls}} \defeq \frac{1}{2^{R_A+r_A+R_B+r_B}}\times\sum_{j=1,k=1}^{\rApls,\rBpls} \Psi_{RABM_jN_k}\otimes \sigma^{(-j)}\otimes \omega^{(-k)} \\  &&
\xi_{RABM_1\ldots M_{\rApls}N_1\ldots N_{\rBpls}}  \defeq\Psi_{RAB}\otimes \sigma_{M_1}\ldots \otimes \sigma_{M_{\rApls}}\otimes \omega_{N_1}\ldots \otimes \omega_{N_{\rBpls}}.
\end{eqnarray*}
Note that $\Psi_{RAB} = \mu_{RAB}$. Let 
\begin{eqnarray*}
\ket{\theta} &=& \ket{\sigma}_{M'_1M_1}\otimes \ldots \ket{\sigma}_{M'_{\rApls}M_{\rApls}}\otimes \ket{\omega}_{N'_1N_1}\otimes \ldots \ket{\omega}_{N'_{\rApls}N_{\rApls}}
\end{eqnarray*}
 be a purification of  $ \sigma_{M_1}\ldots \otimes \sigma_{M_{\rApls}}\otimes \omega_{N_1}\ldots \otimes \omega_{N_{\rBpls}}$. Let 
\begin{eqnarray*}
 &&\ket{\xi} \defeq  \ket{\Psi}_{RABCMN} \otimes\ket{\theta}_{M'_1\ldots M'_{\rApls}N'_1\ldots N'_{\rBpls}M_1\ldots M_{\rApls}N_1\ldots N_{\rBpls}}
\end{eqnarray*}
 be a purification of $\xi_{RABM_1\ldots M_{\rApls}N_1\ldots N_{\rBpls}}$.

Consider the following purification of $\mu_{RAB M_1 \ldots M_{\rApls} N_1 \ldots N_{\rBpls}}$, 
\begin{align*}
&\frac{1}{\sqrt{\rApls\cdot \rBpls}}\sum_{j=1,k=1}^{\rApls,\rBpls}\ket{j,k}_{JK}\otimes\ket{\Psi}_{RABCM_jN_k}\otimes \ket{\sigma^{(-j)}}\otimes\ket{0}_{M'_j}\otimes\ket{\omega^{(-k)}}\otimes\ket{0}_{N'_k},
\end{align*} 
where 
\begin{eqnarray*}
\ket{\sigma^{(-j)}}&\defeq& \ket{\sigma}_{M'_1M_1}\otimes \ldots\otimes\ket{\sigma}_{M'_{j-1}M_{j-1}}\otimes\ket{\sigma}_{M'_{j+1}M_{j+1}}\otimes\ldots\otimes
\ket{\sigma}_{M'_{\rApls}M_{\rApls}}
\end{eqnarray*}
 and 
\begin{eqnarray*}
\ket{\omega^{(-k)}}&\defeq& \ket{\omega}_{N'_1N_1}\otimes \ldots\otimes\ket{\omega}_{N'_{k-1}N_{k-1}}\otimes\ket{\omega}_{N'_{k+1}N_{k+1}} \otimes\ldots\otimes\ket{\omega}_{N'_{\rBpls}N_{\rBpls}} 
\end{eqnarray*}
and $\forall j\in[\rApls]: \ket{\sigma}_{M'_jM_j}$ is a purification of  $\sigma_{M_j}$ and $\forall k\in[\rBpls]: \ket{\omega}_{N'_kN_k}$ is a purification of  $\omega_{N_k}$.

We decompose the register $J$ into registers $J_1,J_2$ satisfying $|J_1|=2^{R_A}, |J_2|=2^{r_A}$. Similarly, we decompose the register $K$ into registers $K_1,K_2$ satisfying $|K_1|=2^{R_B}, |K_2|=2^{r_B}$. Using this, we obtain the following state as a purification of $\mu_{RAB M_1 \ldots M_{\rApls} N_1 \ldots N_{\rBpls}}$ on registers $RABCJ_1J_2K_1K_2M'_1 \ldots M'_{\rApls} N'_1 \ldots N'_{\rBpls}$ $M_1 \ldots M_{\rApls} N_1 \ldots N_{\rBpls}$:

\begin{align*}
&\ket{\mu}\defeq \frac{1}{\sqrt{\rApls\cdot\rBpls}} \sum_{j_1,j_2,k_1,k_2}\ket{j_1,j_2,k_1,k_2}_{J_1J_2K_1K_2}\otimes\ket{\Psi}_{RABCM_jN_k} \otimes \ket{\sigma^{(-j)}}\otimes\ket{0}_{M'_j}\otimes\ket{\omega^{(-k)}}\otimes\ket{0}_{N'_k},
\end{align*} 
where, $j_1\in [1:2^{R_A}], j_2\in [1:2^{r_A}], k_1\in [1:2^{R_B}], k_2\in [1:2^{r_B}]$, $j\defeq (j_1-1)2^{r_A} + j_2$, $k\defeq (k_1-1)2^{r_B} + k_2$. Henceforth, we shall take the convention that $j=(j_1-1)2^{r_A} + j_2, k= (k_1-1)2^{r_B} + k_2$, whenever it is clear from the context.

Using the tripartite convex-split lemma (Lemma ~\ref{triconvexcomb}) and choice of $R_A+r_A,R_B+r_B$, we have
\begin{eqnarray*}
&&\F^2(\xi_{RABM_1\ldots M_{\rApls}N_1\ldots N_{\rBpls}}, \mu_{RABM_1\ldots M_{\rApls}N_1\ldots N_{\rBpls}}) \geq 1 - (\eps_1+2\sqrt{\delta})^2 .
\end{eqnarray*} 
Let $\ket{\xi'}$ be a purification of $\xi_{RABM_1\ldots M_{\rApls}N_1\ldots N_{\rBpls}}$ (guaranteed by Uhlmann's theorem, Fact~\ref{uhlmann}) such that,
\begin{eqnarray}
\label{eq:convclose}
\F^2(\ketbra{\xi'},\ketbra{\mu}) &=&\F^2(\xi_{RABM_1\ldots M_{\rApls}N_1\ldots N_{\rBpls}}, \mu_{RABM_1\ldots M_{\rApls}N_1\ldots N_{\rBpls}})  \nonumber\\ &\geq& 1- (\eps_1+2\sqrt{\delta})^2.
\end{eqnarray} 
Let $V': CMNM'_1\ldots M'_{\rApls}N'_1\ldots N'_{\rBpls} \rightarrow J_1J_2K_1K_2CM'_1\ldots M'_{\rApls}N'_1\ldots N'_{\rBpls}$ be an isometry (guaranteed by Uhlmann's theorem, Fact~\ref{uhlmann})  such that,
$$V'\ket{\xi} = \ket{\xi'} .$$

\vspace{0.1in}

\noindent{\bf 2. The protocol:} Consider the following protocol $\cP$:
\begin{enumerate}
\item Alice, Bob, Charlie and Reference start by sharing the state $\ket{\xi}$ between themselves where Alice holds registers $AM_1 \ldots M_{\rApls}$, Bob holds the registers $BN_1\ldots N_{\rBpls}$, Charlie holds the registers \\ $CMNM'_1\ldots M'_{\rApls}N'_1\ldots N'_{\rBpls}$ and Reference holds the register $R$. Note that  $\ket{\Psi}_{RABCMN}$ is provided as input to the protocol and $\ket{\theta}$ is the additional shared entanglement of the form $\ket{\theta} = \ket{\theta_1}\otimes \ket{\theta_2}$, with $\ket{\theta_1}$ shared between (Alice, Charlie) in registers $E_{AC} \defeq M_1M'_1\ldots M_{\rApls} M'_{\rApls}$ and $\ket{\theta_2}$ shared between (Bob, Charlie) in registers $E_{BC} \defeq N_1N'_1\ldots N_{\rBpls} N'_{\rBpls}$.
\item Charlie applies the isometry $V'$ to obtain state the $\ket{\xi'}$.
\begin{itemize}
\item At this stage of the protocol, the global state $\ket{\xi'}$ is close to the state $\ket{\mu}$, where Alice holds the registers $AM_1\ldots M_{\rApls}$, Bob holds the registers  $BN_1N_2\ldots N_{\rBpls}$, Charlie holds the registers $CJ_1J_2K_1K_2M'_1 \ldots M'_{\rApls} N'_1 \ldots N'_{\rBpls}$ and Reference holds the register $R$. 
\end{itemize}
\item Charlie measures the registers $J_1,K_1$ and obtains the measurement outcomes $(j_1,k_1) \in [1:2^{R_A}]\times [1:2^{R_B}]$. He sends $j_1$ to Alice and $k_1$ to Bob using $\frac{R_A}{2}$ and $\frac{R_B}{2}$ qubits of respective quantum communication. Charlie, Alice and Bob employ coherent superdense coding (\cite{Harrow04, bennett92}) using fresh entanglement to achieve this. Let the final register obtained with Alice be $J'_1$ and with Bob be $K'_1$. 
\begin{itemize}
\item  Note that the additional entanglement for coherent superdense coding is still shared between (Alice, Charlie) and (Bob, Charlie) respectively. Furthermore, the coherent superdense coding scheme does not output any registers other than $J'_1$ and $K'_1$.
\item If global state in step $2$ above were $\ket{\mu}$, the global state at this step would be
\begin{eqnarray*}
\ket{\mu^{(2)}}&\defeq& \frac{1}{\sqrt{\rApls\cdot\rBpls}}\sum_{j_1,j_2,k_1,k_2}\ket{j_1,j_2,k_1,k_2}_{J_1J_2K_1K_2} \otimes\ket{j_1,k_1}_{J'_1,K'_1}\\ &&\otimes\ket{\Psi}_{RABCM_jN_k}\otimes \ket{\sigma^{(-j)}}\otimes\ket{0}_{M'_j}\otimes\ket{\omega^{(-k)}}\otimes\ket{0}_{N'_k}.
\end{eqnarray*} 
\item For a fixed pair $(j_1,k_1)$, we define the state
\begin{align*}
\ket{\mu^{(2)}_{(j_1,k_1)}}&= \frac{1}{\sqrt{2^{r_A+r_B}}}\sum_{j_2,k_2}\ket{j_2,k_2}_{J_2K_2}\otimes\ket{\Psi}_{RABCM_jN_k}\otimes \ket{\sigma^{(-j)}}\otimes\ket{0}_{M'_j}\otimes\ket{\omega^{(-k)}}\otimes\ket{0}_{N'_k}.
\end{align*} 
\end{itemize}

\item This is the hypothesis testing step. Conditioned on $(j_1,k_1)$, Alice and Bob consider the following operations. Let $$\Pi^A_{j_2}\defeq \Pi^A_{AM_{j}}\otimes \id_{M_1}\otimes\ldots \id_{M_{j-1}}\otimes \id_{M_{j+1}}\ldots \otimes \id_{M_{\rApls}},$$ with $j=(j_1-1)2^{r_A} + j_2$ and $\Pi^A\defeq \sum_{j_2} \Pi^A_{j_2}$. Let $$\Pi^B_{k_2}\defeq \Pi^B_{BN_k}\otimes \id_{M_1}\otimes\ldots \id_{M_{k-1}}\otimes \id_{M_{k+1}}\ldots \otimes \id_{M_{\rBpls}},$$ with $k=(k_1-1)2^{r_B} + k_2$ and $\Pi^B\defeq \sum_{k_2}\Pi^B_{k_2}$. Let $(\Pi^A)^0$, which is the operator $\Pi^A$ raised to the power $0$, represent the support of $\Pi^A$. Similarly let $(\Pi^B)^0$ represent the support of $\Pi^B$.

 Alice applies the isometry $\sum_{j_1} \ketbra{j_1}_{J'_1}\otimes \cA_{j_1}$, where 
\begin{eqnarray*}
\cA_{j_1} &\defeq& \sum_{j_2} \sqrt{(\Pi^A)^{-\frac{1}{2}}\Pi^A_{j_2}(\Pi^A)^{-\frac{1}{2}}}\otimes \ket{j_2}_{J'_2} + \sqrt{\id - (\Pi^A)^0}\otimes \ket{0}_{J'_2}.
\end{eqnarray*}
Bob applies the isometry $\sum_{k_1} \ketbra{k_1}_{K'_1}\otimes \cB_{k_1}$, where 
\begin{eqnarray*}
\cB_{k_1} &\defeq& \sum_{k_2} \sqrt{(\Pi^B)^{-\frac{1}{2}}\Pi^B_{k_2}(\Pi^B)^{-\frac{1}{2}}}\otimes \ket{k_2}_{K'_2} + \sqrt{\id - (\Pi^B)^0}\otimes \ket{0}_{K'_2}.
\end{eqnarray*}
Above operations are coherent versions of the position-based decoding operation \cite{AnshuJW17CC} and the outcome $\ket{0}$ corresponds to Alice or Bob not being able to decode any location.  Define 
$$\ket{\mu^{(3)}_{(j_1,k_1)}} : = \cA_{j_1}\otimes \cB_{k_1}\ket{\mu^{(2)}_{(j_1,k_1)}}.$$

\begin{itemize}
\item If the global state on Step $2$ above were $\ket{\mu}$, the resulting global state at this step would be
\begin{eqnarray*}
\ket{\mu^{(3)}} &=& \frac{1}{\sqrt{2^{R_A+R_B}}}\sum_{j_1,k_1}\ket{j_1,k_1}_{J_1K_1}\otimes\ket{j_1,k_1}_{J'_1K'_1}\otimes\ket{\mu^{(3)}_{(j_1,k_1)}}.
\end{eqnarray*}
\item Define the state
\begin{eqnarray*}
\ket{\mu^{(4)}} &=& \frac{1}{\sqrt{2^{R_A+R_B}}}\sum_{j_1,k_1}\ket{j_1,k_1}_{J_1K_1}\otimes\ket{j_1,k_1}_{J'_1K'_1}\otimes\ket{\mu^{(4)}_{(j_1,k_1)}},
\end{eqnarray*}
where
\begin{eqnarray*}
\ket{\mu^{(4)}_{(j_1,k_1)}}&\defeq& \frac{1}{\sqrt{2^{r_A+r_B}}}\sum_{j_2,k_2}\ket{j_2, k_2}_{J_2K_2}\otimes\ket{j_2,k_2}_{J'_2K'_2} \otimes\ket{\Psi}_{RABCM_jN_k}\otimes \\ 
&&\ket{\sigma^{(-j)}}\otimes\ket{0}_{M'_j}\otimes\ket{\omega^{(-k)}}\otimes\ket{0}_{N'_k}.
\end{eqnarray*}
\item We shall show in Claim \ref{claimprotp1} that $\ket{\mu^{(3)}}$ is close to $\ket{\mu^{(4)}}$.
\end{itemize}

\item Alice and Bob introduce the registers $M,N$ in the states $\ket{0}_M,\ket{0}_N$ respectively. Alice applies the operation $$\sum_{j_1,j_2}\ketbra{j_1,j_2}_{J'_1J'_2} \otimes \swapp_{M,M_j},$$
where $j= (j_1-1)\cdot 2^{r_A} + j_2$. Similarly, Bob applies the operation
$$\sum_{k_1,k_2}\ketbra{k_1,k_2}_{K'_1K'_2} \otimes \swapp_{N,N_k},$$
where $k= (k_1-1)\cdot 2^{r_B} + k_2$. 

\begin{itemize}
\item If the global state after step $4$ were $\ket{\mu^{(4)}}$, the resulting global state at this step would be 
\begin{eqnarray*}
\ket{\mu^{(5)}}&\defeq& \frac{1}{\sqrt{\rApls\cdot\rBpls}}\sum_{j_1,j_2,k_1,k_2}\ket{j_1,j_2,k_1,k_2}_{J_1J_2K_1K_2}\otimes\ket{j_1,k_1,j_2,k_2}_{J'_1K'_1J'_2K'_2}  \\ 
&&\otimes\ket{\Psi}_{RAMBNC}\otimes \ket{\sigma^{(-j)}}\otimes\ket{0,0}_{M'_jM_j}\otimes\ket{\omega^{(-k)}}\otimes\ket{0,0}_{N'_k,N_k}\\ &=& \ket{\Psi}_{RAMBNC} \otimes \bigg(\frac{1}{\sqrt{\rApls}}\sum_{j_1,j_2}\ket{j_1,j_2}_{J_1J_2}\ket{j_1,j_2}_{J'_1J'_2} \otimes\ket{\sigma^{(-j)}}\otimes\ket{0,0}_{M'_jM_j}\bigg) \\ && \otimes \bigg(\frac{1}{\sqrt{\rBpls}}\sum_{k_1,k_2}\ket{k_1,k_2}_{K_1K_2}\ket{k_1,k_2}_{K'_1K'_2}\otimes \ket{\omega^{(-k)}}\otimes\ket{0,0}_{N'_kN_k}\bigg),
\end{eqnarray*}
which is of the form $\ket{\Psi}_{RAMBNC} \otimes \ket{\theta'_1}_{E'_{AC}}\otimes \ket{\theta'_2}_{E'_{BC}}$. 
\end{itemize}

\end{enumerate}   

Let $E'_{AC}$ represent the registers $J_1J_2J'_1J'_2M_1M'_1\ldots M_{\rApls} M'_{\rApls}$ and $E'_{BC}$ represent the registers $K_1K_2K'_1K'_2 N_1N'_1\ldots N_{\rBpls} N'_{\rBpls}$. Let $\ketbra{\Phi}_{RAMBNCE'_{AC}E'_{BC}}$ be the output of the protocol $\cP$. 

\vspace{0.1in}

\noindent {\bf 3. Analysis of the protocol:} We shall show that 
$$\Pur(\mu^{(5)}, \Phi_{RAMBNCE'_{AC}E'_{BC}}) \leq \eps_1+2\sqrt{\delta} + 5\eps_2. $$
For this, we shall use the following relations: $\Pur(\ketbra{\xi'},\ketbra{\mu}) \leq \eps_1 + 2\sqrt{\delta}$ (Equation \ref{eq:convclose}) and $\Pur(\mu^{(3)},\mu^{(4)}) \leq 5\eps_2$ (to be shown in Claim \ref{claimprotp1} below).

Let $\ket{\mu^{(6)}}$ be the global quantum state obtained after the application of Step $5$ of the protocol $\cP$ on state $\ket{\mu^{(3)}}$. Thus, $\ket{\mu^{(6)}}$ is also the output obtained after the application of Steps $3-5$ on the state $\ket{\mu}$. Using the monotonicity of fidelity under quantum operation (Fact~\ref{fact:monotonequantumoperation}), we have $$\Pur(\mu^{(6)}, \mu^{(5)}) \leq \Pur(\mu^{(3)}, \mu^{(4)}) \leq 5\eps_2.$$ Moreover, by another application of monotonicity of fidelity under quantum operation (Fact~\ref{fact:monotonequantumoperation}) and the observation that $\ket{\mu^{(6)}}$ is the output of the action of Steps $3-5$ on the state $\ket{\mu}$ , we obtain $$\Pur(\mu^{(6)},\Phi_{RAMBNCE'_{AC}E'_{BC}}) \leq \Pur(\ketbra{\mu}, \ketbra{\xi'}) \leq \eps_1+ 2\sqrt{\delta}.$$ Thus, by the triangle inequality for purified distance (Fact \ref{fact:trianglepurified})  we conclude that
\begin{eqnarray*}
&&\Pur(\mu^{(5)}, \Phi_{RAMBNCE'_{AC}E'_{BC}}) \leq  \Pur(\ketbra{\xi'},\ketbra{\mu}) + \Pur(\mu^{(3)},\mu^{(4)}) \leq \eps_1+2\sqrt{\delta} + 5\eps_2. 
\end{eqnarray*}
This is equivalent to the statement 
\begin{eqnarray*}
&&\Pur(\ketbra{\Psi}_{RAMBNC}\otimes \ketbra{\theta'_1}_{E'_{AC}}\otimes \ketbra{\theta'_2}_{E'_{BC}}, \Phi_{RAMBNCE'_{AC}E'_{BC}}) \leq \eps_1+2\sqrt{\delta} + 5\eps_2.
\end{eqnarray*}
The number of qubits communicated by Charlie to Alice and Charlie to Bob in $\cP$ is $\frac{R_A}{2}$ and $\frac{R_B}{2}$ respectively.

\vspace{0.1in}

\noindent {\bf 4. Hypothesis testing succeeds with high probability:} We show the following claim.

\begin{claim}
\label{claimprotp1}
 It holds that $\F^2(\mu^{(3)},\mu^{(4)}) \geq 1- 24\eps^2_2$.
\end{claim}
\begin{proof}
We shall prove that for every $(j_1,k_1)$, it holds that $\F^2(\mu^{(3)}_{(j_1,k_1)},\mu^{(4)}_{(j_1,k_1)}) \geq 1- 12\eps^2_2$, from which the claim is immediate. Appealing to symmetry, it is sufficient to consider the case $(j_1,k_1)=(1,1)$, for which $j=j_2$ and $k=k_2$. Let $p_{\tj,\tk|j_2,k_2}$ be defined as follows:
\begin{eqnarray*}
p_{\tj,\tk|j_2,k_2} &\defeq& \Tr\bigg((\Pi^A)^{-\frac{1}{2}}\Pi^A_{\tj}(\Pi^A)^{-\frac{1}{2}}\otimes (\Pi^B)^{-\frac{1}{2}}\Pi^B_{\tk}(\Pi^B)^{-\frac{1}{2}}\ketbra{\Psi}_{ABM_{j_2}N_{k_2}}\otimes \sigma^{-j_2}\otimes \omega^{-k_2}\bigg),
\end{eqnarray*}
if $\tj, \tk \neq 0$, 
\begin{eqnarray*}
p_{\tj,0|j_2,k_2} &\defeq& \Tr\bigg((\Pi^A)^{-\frac{1}{2}}\Pi^A_{\tj}(\Pi^A)^{-\frac{1}{2}}\otimes \left(\id - (\Pi^B)^0\right)\ketbra{\Psi}_{ABM_{j_2}N_{k_2}}\otimes \sigma^{-j_2}\otimes \omega^{-k_2}\bigg),
\end{eqnarray*}
if $\tj \neq 0$,
\begin{eqnarray*}
p_{0,\tk|j_2,k_2} &\defeq& \Tr\bigg(\left(\id - (\Pi^A)^0\right)\otimes (\Pi^B)^{-\frac{1}{2}}\Pi^B_{\tk}(\Pi^B)^{-\frac{1}{2}}\ketbra{\Psi}_{ABM_{j_2}N_{k_2}}\otimes\sigma^{-j_2}\otimes \omega^{-k_2}\bigg),
\end{eqnarray*}
if $\tk \neq 0$ and 
\begin{eqnarray*}
p_{0,0|j_2,k_2} &\defeq& \Tr\bigg(\left(\id - (\Pi^A)^0\right)\otimes \left(\id - (\Pi^B)^0\right)\ketbra{\Psi}_{ABM_{j_2}N_{k_2}}\otimes \sigma^{-j_2}\otimes \omega^{-k_2}\bigg).
\end{eqnarray*}

From Lemma \ref{gentlepovm}, it holds that $$\F(\cA_1\otimes \cB_1(\mu^{(2)}_{(1,1)}),\mu^{(4)}_{(1,1)})\geq \frac{1}{2^{r_A+r_B}}\sum_{j_2,k_2} p_{j_2,k_2|j_2,k_2}.$$ Since $$\frac{1}{2^{r_A+r_B}}\sum_{j_2,k_2,\tj,\tk}p_{\tj,\tk|j_2,k_2} = 1,$$ we have that 
\begin{eqnarray*}
\frac{1}{2^{r_A+r_B}}\sum_{j_2,k_2} p_{j_2,k_2|j_2,k_2}  &=& 1-\frac{1}{2^{r_A+r_B}}\sum_{j_2,k_2}\sum_{\tj\neq j_2, \tk\neq k_2} p_{\tj,\tk|j_2,k_2}  \\ &=& 1-\sum_{\tj\neq 1, \tk\neq 1} p_{\tj,\tk|1,1},
\end{eqnarray*}
where the last line follows by symmetry under interchange of registers $M_{j_2},N_{k_2}$. Now, consider

\begin{eqnarray*}
\sum_{\tj\neq 1,\tk\neq 1}p_{\tj,\tk|1,1} &=& \Tr\bigg(\bigg(\id_A\otimes \id_B - (\Pi^A)^{-\frac{1}{2}}\Pi^A_{1}(\Pi^A)^{-\frac{1}{2}}\otimes (\Pi^B)^{-\frac{1}{2}}\Pi^B_{1}(\Pi^B)^{-\frac{1}{2}}\bigg) \ketbra{\Psi}_{ABM_1N_1}\otimes \sigma^{-1}\otimes \omega^{-1}\bigg)\\  &\overset{a}\leq& \Tr\bigg(\left(\id_A - (\Pi^A)^{-\frac{1}{2}}\Pi^A_{1}(\Pi^A)^{-\frac{1}{2}}\right)\otimes \id_B\cdot \ketbra{\Psi}_{ABM_1N_1}\otimes \sigma^{-1}\otimes \omega^{-1}\bigg) \\ &&+ \Tr\bigg(\id_A\otimes \left(\id_B- (\Pi^B)^{-\frac{1}{2}}\Pi^B_{1}(\Pi^B)^{-\frac{1}{2}}\right) \cdot \ketbra{\Psi}_{ABM_1N_1}\otimes \sigma^{-1}\otimes \omega^{-1}\bigg)\\ &\overset{b}\leq& \Tr\bigg(\left(2(\id_A - \Pi_1^A) + 4\sum_{\tj\neq 1}\Pi_{\tj}^A\right)\ketbra{\Psi}_{AM_1}\otimes \sigma^{-1}\bigg) \\ && + \Tr\bigg(\bigg(2(\id_B - \Pi_1^B) + 4\sum_{\tk\neq 1}\Pi_{\tk}^B\bigg) \ketbra{\Psi}_{BN_1}\otimes  \omega^{-1}\bigg)\\ &\overset{c}\leq&  4(\eps^2_2) +  4\cdot 2^{r_A-\dmineps{\Psi_{AM}}{\Psi_A\otimes \sigma_M}{\eps_2^2}} + 4\cdot 2^{r_B-\dmineps{\Psi_{BN}}{\Psi_B\otimes \omega_N}{\eps_2^2}}\leq 12\eps_2^2,
\end{eqnarray*}

where in (a) we use the operator inequality $$\left(\id- P\otimes Q\right) \preceq \id\otimes ( \id - Q) + (\id - P)\otimes \id,$$ for positive semidefinite operators $P,Q \preceq \id$ ; (b) follows from the Hayashi-Nagaoka inequality (Fact \ref{haynag}), (c) follows from the definition of $\Pi^A,\Pi^B$ and the choice of $r_A,r_B$. This implies that $\frac{1}{2^{r_A+r_B}}\sum_{j_2,k_2} p_{j_2,k_2|j_2,k_2} \geq 1-12\eps^2_2$. 

Thus, 
\begin{eqnarray*}
\F^2(\mu^{(3)}_{(1,1)}, \mu^{(4)}_{(1,1)}) = \F^2(\cA_1\otimes \cB_1(\mu^{(2)}_{(1,1)}),\mu^{(4)}_{(1,1)}) \geq (1-12\eps^2_2)^2 \geq 1-24\eps^2_2,
\end{eqnarray*} 
from which the claim concludes.

\end{proof}

This completes the proof of the theorem.

\end{proof}

\subsection*{Achievable rate region for Task $1$}
\suppress{
\begin{figure}[ht]
\centering
\begin{tikzpicture}[xscale=0.9,yscale=1.1]
%\draw[help lines] (0,0) grid (8,6);

\draw[ultra thick] (0,6) -- (0,0) -- (8,0);
\node at (-0.7, 5.8) {$R_{A\to C}$};
\node at (7.8,-0.3){$R_{B\to C}$};

\draw[thick] (1,6) -- (1,3) -- (3,1) -- (8,1);

\draw[thick] (0.6,6) -- (0.6,2.8) -- (2,1.4) -- (8,1.4);

\draw[thick] (1.4,6) -- (1.4,2.25) -- (3.05,0.6) -- (8,0.6);

\draw [thick, gray](2.5,4.5) -- (3.5,5.5);
\draw [thick, gray](3,4) -- (4,5);
\draw [thick, gray](3.5,3.5) -- (4.5,4.5);
\draw [thick, gray](4,3) -- (5,4);
\draw [thick, gray](4.5,2.5) -- (5.5,3.5);
\draw [thick, gray](5,2) -- (6,3);

\end{tikzpicture}
\caption{\small The achievable rate region obtained in Theorem \ref{thm:task1} is the union of rate regions specified by a collection of states $\Psi' \in \ball{\eps}{\Psi}$ that satisfy $\Psi'_{RAB} \preceq 2^{\delta}\Psi_{RAB}$. The black lines represent the inner boundary of the rate region for three different choices of states $\Psi'$.  }
 \label{fig:rateregion}
\end{figure}
}

\begin{theorem}
\label{thm:task1}
Fix $\eps_1,\eps_2,\delta\in (0,1)$ such that $\eps_1+5\eps_2+2\sqrt{\delta} <1$. Let Alice $(AM)$, Bob $(BN)$, Reference $(R)$ and Charlie $(C)$ share the pure state $\ket{\Psi}_{RAMBNC}$. There exists an entanglement assisted quantum protocol, with entanglement shared only between (Alice, Charlie) and (Bob, Charlie), such that at the end of the protocol, Alice $(A)$, Bob $(B)$, Reference $(R)$ and Charlie $(CMN)$ share the state $\Phi'_{RABCMN}$ with the property that $\Pur(\Phi',\ketbra{\Psi})\leq \eps_1+5\eps_2+2\sqrt{\delta}$. The number of qubits that Alice sends to Charlie is $R_{A\to C}$ and that Bob sends to Charlie is $R_{B\to C}$, where the pair $(R_{A\to C},R_{B\to C})$ lie in the union of the following achievable rate region: for every $\Psi'_{RABCMN} \in \ball{\eps_1}{\Psi_{RABCMN}}$ such that $\Psi'_{RAB} \preceq 2^{\delta}\Psi_{RAB}$ and states $\sigma_M,\omega_N$:

\begin{eqnarray*}
R_{A\to C} &\geq& \frac{1}{2}\bigg(\dmax{\Psi'_{RABM}}{\Psi_{RAB}\otimes \sigma_M} - \dmineps{\Psi_{AM}}{\Psi_A\otimes \sigma_M}{\eps_2^2} + \log\frac{1}{\eps_2^2\delta}\bigg),\\
R_{B\to C} &\geq& \frac{1}{2}\bigg(\dmax{\Psi'_{RABN}}{\Psi_{RAB}\otimes \omega_N} - \dmineps{\Psi_{BN}}{\Psi_B\otimes \omega_N}{\eps_2^2} + \log\frac{1}{\eps_2^2\delta}\bigg), \\
R_{A\to C}+R_{B\to C}&\geq& \frac{1}{2}\bigg(\dmax{\Psi'_{RABMN}}{\Psi_{RAB}\otimes\sigma_M\otimes\omega_N} -\dmineps{\Psi_{AM}}{\Psi_A\otimes \sigma_M}{\eps_2^2}\\ &&-\dmineps{\Psi_{BN}}{\Psi_B\otimes \omega_N}{\eps_2^2} +\log\frac{1}{\eps_2^2\delta}\bigg) .
\end{eqnarray*}  
\end{theorem}

\begin{proof}
Consider the protocol $\cP$ as obtained in Theorem \ref{thm:task2} for the Task $2$, with the starting state $\ketbra{\Psi}_{RAMBNC}\otimes \ketbra{\theta_1}_{E_{AC}}\otimes \ketbra{\theta_2}_{E_{BC}}$ (where $\theta_1$ and $\theta_2$ serve as pre-shared entanglement) and the final state $\ketbra{\Phi}_{RABCMNE'_{AC}E'_{BC}}$. Furthermore, as promised by Theorem \ref{thm:task2}, there exist states $\ket{\theta'_1}_{E'_{AC}}$ and $\ket{\theta'_2}_{E'_{BC}}$ such that $$\Pur(\ketbra{\Phi},\ketbra{\Psi}\otimes\ketbra{\theta'_1}\otimes\ketbra{\theta'_2})\leq \eps_1+5\eps_2+2\sqrt{\delta}.$$ Since the protocol can be viewed as a unitary by Charlie, followed by quantum communication from Charlie to Alice and Bob and then subsequent unitaries by Alice and Bob, this protocol can be reversed to obtain a protocol $\cP'$. We take $\cP'$ as the desired protocol for above task and let $\Phi'_{RABCMNE_{AC}E_{BC}}$ be the state obtained by running $\cP'$ on $\ketbra{\Psi}\otimes\ketbra{\theta'_1}\otimes\ketbra{\theta'_2}$ (where $\ket{\Psi}$ serves as input to the protocol, $\ket{\theta'_1}$ serves as the shared entanglement between (Alice, Charlie) and $\ket{\theta'_2}$ serves as the shared entanglement between (Bob,Charlie)). From the relation $\cP'(\ketbra{\Phi}) = \ketbra{\Psi}\otimes \ketbra{\theta_1}\otimes \ketbra{\theta_2}$ and the monotonicity of fidelity under quantum operations (Fact \ref{fact:monotonequantumoperation}), we conclude 
\begin{eqnarray*}
\Pur(\ketbra{\Psi}\otimes \ketbra{\theta_1}\otimes \ketbra{\theta_2}, \Phi') &=& \Pur(\cP'(\ketbra{\Phi}), \cP'(\ketbra{\Psi}\otimes\ketbra{\theta'_1}\otimes\ketbra{\theta'_2}))\\ &\leq & \Pur(\ketbra{\Phi}, \ketbra{\Psi}\otimes\ketbra{\theta'_1}\otimes\ketbra{\theta'_2})\\ &\leq& \eps_1+5\eps_2+2\sqrt{\delta}.
\end{eqnarray*}
This completes the proof.
\end{proof}

\section{Achievable rate region in the asymptotic and i.i.d. setting}
\label{sec:asymptoticiid}

In this section, we re-derive the result of \cite{ADHW09}, but without the use of time-sharing.  Consider the asymptotic and i.i.d. version of Task $1$ (Section \ref{sec:main}) in the special case where register $C$ is trivial, i.e., the joint state between Alice, Bob and Reference is the n-fold tensor product of the state $\ket{\Psi}_{RAMBN}$. Using Theorem \ref{thm:task1} (with $\sigma_M\rightarrow \Psi_M, \omega_N\rightarrow \Psi_N$), Theorem \ref{thm:asymptotic} (in Appendix~\ref{sec:asymptote}) for the pure state $\ket{\Psi}_{RAMBN}$ and Fact \ref{dmaxequi}, we conclude that the rate pair $(R_{A\to C},R_{B \to C})$ is asymptotic and i.i.d. achievable for Task $1$, if it satisfies the following constraints:
\begin{eqnarray*}
R_{A\to C} &\geq& \frac{1}{2}\left(\mutinf{RAB}{M}_{\Psi} - \mutinf{A}{M}_{\Psi} \right),\\
R_{B\to C} &\geq& \frac{1}{2}\left(\mutinf{RAB}{N}_{\Psi} - \mutinf{B}{N}_{\Psi} \right), \\
R_{A\to C}+R_{B\to C}&\geq& \frac{1}{2}\bigg(\trimutinf{RAB}{M}{N}_{\Psi}-\mutinf{A}{M}_{\Psi}-\mutinf{B}{N}_{\Psi}\bigg).
\end{eqnarray*}  

\noindent {\bf Quantum version of the achievable rate region obtained by Slepian and Wolf:} An immediate corollary of above achievable rate region is the following. Consider the task in \cite{ADHW09}, which is a quantum version of the Slepian-Wolf protocol \cite{SlepianW73}. Alice $(M^n)$, Bob $(N^n)$, Reference $(R^n)$ share the joint pure state $\ket{\Psi}^{\otimes n}_{RMN}$.  Alice and Bob wish to communicate their registers $M^n$ and $N^n$ to Charlie such that the final state shared between Reference $(R^n)$ and Charlie $(M^nN^n)$ is $\Phi_{R^nM^nN^n}$ such that $\lim_{n\rightarrow \infty}\Pur(\Psi_{RMN}^{\otimes n},\Phi_{R^nM^nN^n}) = 0$. To accomplish this task, there exists an entanglement assisted protocol (with the entanglement shared between (Alice, Charlie) and (Bob, Charlie)) if the amount of communication from Alice to Charlie ($R_{A\to C}$) and Bob to Charlie ($R_{B\to C}$) satisfy the following constraints  
\begin{eqnarray*}
R_{A\to C} &\geq& \frac{1}{2}\mutinf{R}{M}_{\Psi},\\
R_{B\to C} &\geq& \frac{1}{2}\mutinf{R}{N}_{\Psi}, \\
R_{A\to C}+R_{B\to C}&\geq& \frac{1}{2}\trimutinf{R}{M}{N}_{\Psi}.
\end{eqnarray*}  

\section{Converse bound}
\label{sec:converse}

In this section, we establish a converse for Task $2$ in the absence of registers $A,B$. This matches with the achievable rate region in Theorem \ref{thm:task2}.  We show the following theorem.

\begin{theorem}
\label{theo:conversetask2}
Fix $\eps \in (0,1)$. Let Reference $(R)$ and Charlie $(CMN)$ share the pure state $\ket{\Psi}_{RCMN}$. Let $\cP$ be an entanglement-assisted quantum protocol with the following properties. 
\begin{itemize}
\item The pre-shared entanglement is of the form $\ket{\theta_1}_{E_AE_{C_A}}\otimes\ket{\theta_2}_{E_BE_{C_B}}$, where $\ket{\theta_1}_{E_AE_{C_A}}$ is shared between Alice ($E_A$), Charlie ($E_{C_A}$) and $\ket{\theta_2}_{E_BE_{C_B}}$ is shared between Bob ($E_B$), Charlie ($E_{C_B}$).
\item  The quantum communication is from Charlie to Alice with $R_{C\to A}$ qubits and from Charlie to Bob with $R_{C\to B}$ qubits. \item At the end of the protocol, the joint quantum state $\Phi_{RMNC}$ shared between Alice (M), Bob (N), Reference (R) and Charlie (C)  satisfies $\Phi_{RMNC}\in \ball{\eps}{\Psi_{RMNC}}$.
\end{itemize}
Then there exists a quantum state $\Psi'_{RMNC} \in \ball{\eps}{\Psi_{RMNC}}$ with $\Psi'_R = \Psi_R$ and quantum states $\sigma_M, \omega_N$ such that 
\begin{eqnarray*}
R_{C\to A} &\geq& \frac{1}{2}\dmax{\Psi'_{RM}}{\Psi_{R}\otimes \sigma_M},\\
R_{C\to B} &\geq& \frac{1}{2}\dmax{\Psi'_{RN}}{\Psi_{R}\otimes \omega_N}, \\
R_{C\to A}+R_{C\to B}&\geq& \frac{1}{2}\dmax{\Psi'_{RMN}}{\Psi_{R}\otimes\sigma_M\otimes\omega_N}.\\
\end{eqnarray*}  
\end{theorem}
\begin{proof}
A protocol $\cP$ has the following steps, where the registers $Q_A, Q_B$ serve as the message registers from Charlie to Alice and Bob, respectively.
\begin{itemize}
\item Charlie applies an encoding map $\cE: CMNE_{C_A}E_{C_B} \to CQ_AQ_B$ and communicates $Q_A$ and $Q_B$ to Alice and Bob, respectively. 
\item Alice applies a decoding map $\cD_A: E_AQ_A \rightarrow M$ and Bob applies a decoding map $\cD_B: E_BQ_B \rightarrow N$.
\item The final quantum state is obtained in the registers $RCMN$.
\end{itemize}

Let the quantum state on the registers $RQ_AE_AQ_BE_B$ after Alice and Bob receive Charlie's message be $\Omega_{RQ_AE_AQ_BE_B}$. Observe that $\Omega_{RE_AE_B} = \Psi_R\otimes (\theta_1)_{E_A}\otimes (\theta_2)_{E_B}$ and the final state $\Phi_{RMNC}$ is equal to $\cD_A\otimes \cD_B (\Omega_{RQ_AE_AQ_BE_B})$.
We have the following relations using Fact \ref{fact:dmaxup},
\begin{eqnarray}
\label{conv:dmaxlower}
\log|Q_A| &\geq& \frac{1}{2}\dmax{\Omega_{RQ_AE_A}}{\Omega_{RE_A}\otimes \frac{\id_{Q_A}}{|Q_A|}} = \frac{1}{2}\dmax{\Omega_{RQ_AE_A}}{\Psi_R\otimes (\theta_1)_{E_A}\otimes \frac{\id_{Q_A}}{|Q_A|}} ,\nonumber\\
\log|Q_B| &\geq& \frac{1}{2}\dmax{\Omega_{RQ_BE_B}}{\Omega_{RE_B}\otimes \frac{\id_{Q_B}}{|Q_B|}} = \frac{1}{2}\dmax{\Omega_{RQ_BE_B}}{\Psi_R\otimes (\theta_2)_{E_B}\otimes \frac{\id_{Q_B}}{|Q_B|}} ,\nonumber\\
\log(|Q_A|\cdot |Q_B|) &\geq& \frac{1}{2}\dmax{\Omega_{RQ_AE_AQ_BE_B}}{\Omega_{RE_AE_B}\otimes \frac{\id_{Q_A}}{|Q_A|}\otimes \frac{\id_{Q_B}}{|Q_B|}} \nonumber\\ &=& \frac{1}{2}\mathrm{D}_{\max}\bigg(\Omega_{RQ_AE_AQ_BE_B}\|  \Psi_R\otimes (\theta_1)_{E_A}\otimes (\theta_2)_{E_B}\otimes \frac{\id_{Q_A}}{|Q_A|}\otimes \frac{\id_{Q_B}}{|Q_B|}\bigg).
\end{eqnarray}

Define $\Psi'_{RMNC} \defeq \Phi_{RMNC} = \cD_A\otimes \cD_B (\Omega_{RQ_AE_AQ_BE_B})$. It holds that $\Psi'_R = \Phi_R = \Omega_R = \Psi_R$. Further, define $\sigma_M \defeq \cD_A\left((\theta_1)_{E_A}\otimes \frac{\id_{Q_A}}{|Q_A|}\right)$ and $\omega_N \defeq \cD_B\left((\theta_2)_{E_B}\otimes \frac{\id_{Q_B}}{|Q_B|}\right)$. Applying the monotonicity of max-relative entropy under quantum operations (Fact \ref{fact:monotonequantumoperation}) in Equation \ref{conv:dmaxlower}, we obtain
\begin{eqnarray*}
\log|Q_A| &\geq& \frac{1}{2}\dmax{\cD_A(\Omega_{RQ_AE_A})}{\Psi_R\otimes \cD_A\left((\theta_1)_{E_A}\otimes \frac{\id_{Q_A}}{|Q_A|}\right)}\\ &=& \frac{1}{2}\dmax{\Psi'_{RM}}{\Psi_R\otimes \sigma_M} ,\\
\log|Q_B| &\geq& \frac{1}{2}\dmax{\cD_B(\Omega_{RQ_BE_B})}{\Psi_R\otimes \cD_B\left((\theta_2)_{E_B}\otimes \frac{\id_{Q_B}}{|Q_B|}\right)} \\ &=& \frac{1}{2}\dmax{\Psi'_{RN}}{\Psi_R\otimes \omega_N} ,\\
\log(|Q_A|\cdot |Q_B|) &\geq& \frac{1}{2}\mathrm{D}_{\max}\bigg(\cD_A\otimes \cD_B(\Omega_{RQ_AE_AQ_BE_B})\|\Psi_R\otimes\cD_A\left((\theta_1)_{E_A}\otimes \frac{\id_{Q_A}}{|Q_A|}\right) \otimes \cD_B\left((\theta_2)_{E_B} \otimes \frac{\id_{Q_B}}{|Q_B|}\right)\bigg)\\
&=& \frac{1}{2}\dmax{\Psi'_{RMN}}{\Psi_R\otimes \sigma_M\otimes \omega_N}.
\end{eqnarray*}
Since $R_{C\to A} = \log|Q_A|, R_{C\to B} = \log|Q_B|$, the theorem concludes.
\end{proof}

\subsection*{Conclusion}

In this work, we have studied two distributed quantum source compression tasks characterized by two senders-one receiver and one sender-two receivers, respectively. These cases generalize the distributed source compression tasks studied in \cite{ADHW09} and \cite{HsiehW2015} and the task of quantum state redistribution \cite{Devatakyard, YardD09}. We have obtained one-shot achievable rate regions for these tasks and shown matching converse in a special case. 

An important question for our one-shot achievability results is to connect them to the relative entropy in the asymptotic and i.i.d. setting. We are able to achieve this for a special case of our Task (which is still general enough to include the task considered in \cite{ADHW09}). But it is not clear to us how to achieve the same for our most general setting. In fact, a similar problem arises in attempting to extend our work to more complicated network scenarios. We are able to obtain one-shot achievable rate regions in complicated network scenarios that closely resemble the achievable rate region in Theorem \ref{thm:task2}. However, we are unable to show that such rate regions converge appropriately in the asymptotic and i.i.d. setting. We leave this as an important question to be pursued for future work. 

Finally, we leave unexplored the study of a variant of our tasks where all the parties are allowed to pre-share tri-partite entanglement. Such a scenario might allow us to prove near optimal converse bounds for Task $1$. This would be in striking contrast with the classical analogue studied by Slepian and Wolf \cite{SlepianW73}, where the optimal rate region can be achieved without any shared randomness between the senders.

\subsection*{Acknowledgment} 
We thank the anonymous referees for their helpful suggestions and comments.

This work is supported by the Singapore Ministry of Education and the National Research Foundation,
through the Tier 3 Grant ``Random numbers from quantum processes'' MOE2012-T3-1-009 and NRF RF Award NRF-NRFF2013-13. 

\bibliographystyle{alpha}
\bibliography{References}

\appendix

\section{A variant of the convex-split lemma}
\label{sec:convexsplit}
In this section, we prove a tripartite variant of the convex split lemma, used in the proof of Theorem \ref{thm:task2}. We shall use the following fact.

\begin{fact}[\cite{ADJ14}]
\label{relentconcav}
Let $\mu_1,\mu_2,\ldots \mu_n, \theta$ be quantum states and $\{p_1,p_2,\ldots p_n\}$ be a probability distribution. Let $\mu=\sum_i p_i\mu_i$ be the average quantum state. Then 
$$\relent{\mu}{\theta} =\sum_i p_i(\relent{\mu_i}{\theta}-\relent{\mu_i}{\mu}).$$
\end{fact}

Using this fact, we prove the following statement.

\begin{lemma}[tripartite convex-split lemma]
\label{triconvexcomb}
Let $\eps,\delta \in (0,1)$ such that $\eps+2\sqrt{\delta}<1$. Let $\rho_{RAB}\in\mathcal{D}(RAB), \sigma_A\in \mathcal{D}(A), \omega_B\in \mathcal{D}(B)$ be quantum states and $\rho'_{RAB}$ be a quantum state satisfying $\rho'_{RAB}\in \ball{\eps}{\rho_{RAB}}$. For some $R_1,R_2\geq 1$, consider the following state
\begin{eqnarray*}
&&\tau_{RA_1\ldots A_{\rA} B_1\ldots B_{\rB}} \defeq  \frac{1}{2^{R_A+R_B}}\sum_{i=1}^{\rA}\sum_{j=1}^{\rB} \rho_{RA_iB_j}\\ &&\otimes\sigma_{A_1}\otimes\ldots\sigma_{A_{i-1}}\otimes \sigma_{A_{i+1}}\otimes\ldots\sigma_{A_{\rA}}\\ &&\otimes\omega_{B_1}\otimes \ldots\omega_{B_{j-1}}\otimes\omega_{B_{j+1}}\ldots\otimes\omega_{B_{\rB}}
\end{eqnarray*}
on the registers $A_1,A_2\ldots A_{\rA},B_1,B_2,\ldots B_{\rB}$, where $\forall i \in [\rA],j\in [\rB]: \rho_{RA_iB_j} = \rho_{RAB}$, $\rho_{RA_i}=\rho_{RA}$ and $\rho_{RB_j}=\rho_{RB}$. If 
\begin{eqnarray*}
&& R_A \geq \dmax{\rho'_{RA}}{\rho_R\otimes\sigma_A} + \log\frac{1}{\delta},\\
&& R_B \geq \dmax{\rho'_{RB}}{\rho_R\otimes\omega_B} + \log\frac{1}{\delta}, \\
&& R_A+R_B \geq \dmax{\rho'_{RAB}}{\rho_R\otimes\sigma_A\otimes\omega_B} +\log\frac{1}{\delta},\\
&& \rho'_R \preceq 2^{\delta}\rho_R
\end{eqnarray*}    
then 
\begin{eqnarray*} 
&&\Pur(\tau_{RA_1A_2\ldots A_{\rA}B_1B_2\ldots B_{\rB}},\rho_R\otimes\sigma_{A_1}\otimes\sigma_{A_2}\otimes\ldots\sigma_{A_{\rA}} \\ &&\otimes \omega_{B_1}\otimes\omega_{B_2}\ldots \otimes \omega_{B_{\rB}}) \leq \eps+2\sqrt{\delta}.
\end{eqnarray*}
\end{lemma}

The proof closely follows the original proof of the convex split lemma from~\cite{ADJ14}.
\begin{proof}

For brevity, we set
\begin{eqnarray*}
&& k_1\defeq\dmax{\rho'_{RAB}}{\rho_R\otimes\sigma_A\otimes\omega_B},\\
&& k_2\defeq \dmax{\rho'_{RA}}{\rho_R\otimes\sigma_A},\\
&& k_3\defeq\dmax{\rho'_{RB}}{\rho_R\otimes\omega_B}.
\end{eqnarray*}

 We shall work with the state 
\begin{eqnarray*}
&&\tau'_{RA_1\ldots A_{\rA}B_1\ldots B_{\rB}} \defeq  \frac{1}{\rA\cdot \rB}\sum_{i=1}^{\rA}\sum_{j=1}^{\rB} \rho'_{RA_iB_j}\\ &&\otimes\sigma_{A_1}\otimes\ldots\sigma_{A_{i-1}}\otimes \sigma_{A_{i+1}}\otimes\ldots\sigma_{A_{\rA}}\\ &&\otimes\omega_{B_1}\otimes \ldots\omega_{B_{j-1}}\otimes\omega_{B_{j+1}}\ldots\otimes\omega_{B_{\rB}}.
\end{eqnarray*}

Define, 
\begin{eqnarray*}
\rho^{-(i,j)} &\defeq& \sigma_{A_1}\otimes\ldots\sigma_{A_{i-1}}\otimes \sigma_{A_{i+1}}\otimes\ldots\sigma_{A_{\rA}}\\ &&\otimes\omega_{B_1}\ldots\otimes\omega_{B_{j-1}}\otimes \omega_{B_{j+1}}\ldots \otimes \omega_{B_{\rB}},
\end{eqnarray*} 

$$\rho\defeq \sigma_{A_1}\otimes\sigma_{A_2}\otimes\ldots\sigma_{A_{\rA}}\otimes  \omega_{B_1}\otimes \omega_{B_2}\ldots \omega_{B_{\rB}}. $$ Then 
$$\tau'_{RA_1A_2\ldots A_{\rA}B_1B_2\ldots B_{\rB}}=\frac{1}{\rA\cdot \rB}\sum_{i,j}\rho'_{RA_iB_j}\otimes \rho^{-(i,j)}. $$ 
Now, we use Fact \ref{relentconcav} to express
\begin{eqnarray}
\label{eq:convsplit}
&& \relent{\tau'_{RA_1\ldots A_{\rA}B_1\ldots B_{\rB}}}{\rho_R\otimes \rho} \nonumber\\ &&= \frac{1}{\rA\cdot \rB}\sum_{i,j} \relent{\rho'_{RA_iB_j}\otimes \rho^{-(i,j)}}{\rho_R\otimes\rho} \nonumber\\ &&  - \frac{1}{\rA\cdot \rB}\sum_{i,j}\mathrm{D}\bigg(\rho'_{RA_iB_j}\otimes \nonumber\\ && \hspace{1cm}\rho^{-(i,j)}\|\tau'_{RA_1A_2\ldots A_{\rA}B_1B_2\ldots B_{\rB}}\bigg).
\end{eqnarray}
Note that,
\begin{eqnarray*}
&&\relent{\rho'_{RA_iB_j}\otimes \rho^{-(i,j)}}{\rho_R\otimes \rho} \\ &&= \relent{\rho'_{RA_iB_j}}{\rho_R\otimes\sigma_{A_i}\otimes \omega_{B_j}}, \\ && \relent{\rho'_{RA_iB_j}\otimes \rho^{-(i,j)}}{\tau'_{RA_1A_2\ldots A_{\rA}B_1B_2\ldots B_{\rB}}} \\ &&\geq \relent{\rho'_{RA_iB_j}}{\tau'_{RA_iB_j}},
\end{eqnarray*} 
as relative entropy decreases under partial trace. Moreover, 
\begin{eqnarray*}
&&\tau'_{RA_iB_j} = \frac{1}{\rA\cdot \rB}\rho'_{RA_iB_j} \\ &&+ \frac{1}{\rA}(1-\frac{1}{\rB})\rho'_{RA_i}\otimes \omega_{B_j} \\ &&+ \frac{1}{\rA}(1-\frac{1}{\rB})\sigma_{A_i}\otimes \rho'_{RB_j}\\ &&+ (1-\frac{1}{\rA}-\frac{1}{\rB}+\frac{1}{\rA\cdot \rB})\rho'_R\otimes\sigma_{A_i}\otimes\omega_{B_j}.
\end{eqnarray*}
 By assumption, $$\rho'_{RA_iB_j} \preceq 2^{k_1}\rho_R\otimes\sigma_{A_i}\otimes \omega_{B_j}, \quad \rho'_{RA_i}\preceq 2^{k_2}\rho_R\otimes\sigma_{A_i},$$$$ \rho'_{RB_i}\preceq 2^{k_3}\rho_R\otimes\omega_{B_i}, \quad \rho'_R\preceq 2^{\delta}\rho_R.$$ Hence 
 $$\tau'_{RA_iB_j} \preceq (2^{\delta}+ \frac{2^{k_2}}{\rA}+\frac{2^{k_3}}{\rB}+\frac{2^{k_1}}{\rA\cdot \rB})\rho_R\otimes\sigma_{A_i}\otimes\omega_{B_j}.$$ Since $\log(A)\preceq \log(B)$ if $A \preceq B$, for positive semi-definite matrices $A$ and $B$, we have
\begin{eqnarray*}
&&\relent{\rho'_{RA_iB_j}}{\tau'_{RA_iB_j}} \\ &&= \Tr(\rho'_{RA_iB_j}\log\rho'_{RA_iB_j}) - \Tr(\rho'_{RA_iB_j}\log\tau'_{RA_iB_j}) \\ &&\geq \Tr(\rho'_{RA_iB_j}\log\rho'_{RA_iB_j})  \\ &&- \Tr(\rho'_{RA_iB_j}\log(\rho_R\otimes\sigma_{A_i}\otimes\omega_{B_j})) \\ &&- \log\left(2^{\delta}+ \frac{2^{k_2}}{\rA}+\frac{2^{k_3}}{\rB}+\frac{2^{k_1}}{\rA\cdot \rB}\right) \\ && = \relent{\rho'_{RA_iB_j}}{\rho_R\otimes\sigma_{A_i}\otimes\omega_{B_j}} \\ &&- \log\left(2^{\delta}+ \frac{2^{k_2}}{\rA}+\frac{2^{k_3}}{\rB}+\frac{2^{k_1}}{\rA\cdot \rB}\right) .
\end{eqnarray*}
Using in Equation \ref{eq:convsplit}, we find that
\begin{eqnarray*}
&&\relent{\tau'_{RA_1A_2\ldots A_{\rA}B_1B_2\ldots B_{\rB}}}{\rho_R\otimes\rho}\\ &&\leq \frac{1}{\rA\cdot \rB}\sum_{i,j} \relent{\rho'_{RA_iB_j}}{\rho_R\otimes\sigma_{A_i}\otimes\omega_{B_j}} \\ &&- \frac{1}{\rA\cdot \rB}\sum_{i,j}\relent{\rho'_{RA_iB_j}}{\rho_R\otimes\sigma_{A_i}\otimes\omega_{B_j}}\\ &&+ \log\left(2^{\delta}+ \frac{2^{k_2}}{\rA}+\frac{2^{k_3}}{\rB}+\frac{2^{k_1}}{\rA\cdot \rB}\right) \\ && \leq \log\left(1+\delta+ 3\delta\right).
\end{eqnarray*}
Above, the last inequality follows by the lower bound on $\rA,\rB$ and the fact that $\delta<1$. Thus, by Fact \ref{pinsker} (which is the improved version of Pinsker's inequality), we obtain 
$$\Pur(\tau'_{RA_1A_2\ldots A_{\rA}B_1B_2\ldots B_{\rB}},\rho_R\otimes\rho) \leq \sqrt{4\delta}.$$
Since $\Pur(\tau'_{A_1A_2\ldots A_{\rA}B_1B_2\ldots B_{\rB}},\tau_{A_1A_2\ldots A_{\rA}B_1B_2\ldots B_{\rB}})\leq \Pur(\rho'_{AB},\rho_{AB})\leq \eps$, triangle inequality for Purified distance (Fact \ref{fact:trianglepurified}) shows that 

$$\Pur(\tau_{RA_1A_2\ldots A_{\rA}B_1B_2\ldots B_{\rB}},\rho_R\otimes\rho) \leq \eps+2\sqrt{\delta}.$$
This proves the lemma.
\end{proof}

\section{Asymptotic and i.i.d. analysis}
\label{sec:asymptote}

An important property of the smooth information theoretic quantities is that in the asymptotic and i.i.d. setting, they converge to the relative entropy based quantities. In this section, we show this property for the one-shot bounds we obtain in Theorem \ref{thm:task1}, in the case where register $C$ is absent. 

\subsection*{Facts used in the proof}

Following fact ensures that the smooth max-relative entropy and hypothesis testing relative entropy converge to suitable quantities in the asymptotic and i.i.d. setting.

\begin{fact}[\cite{TomHay13, li2014}]
\label{dmaxequi}
Let $\eps\in (0,1)$ and $n$ be an integer. Let $\rho^{\otimes n}, \sigma^{\otimes n}$ be quantum states. Define $V(\rho\|\sigma) = \Tr(\rho(\log\rho - \log\sigma)^2) - (\relent{\rho}{\sigma})^2$ and $\Phi(x) = \int_{-\infty}^x \frac{e^{-x^2/2}}{\sqrt{2\pi}} dx$. It holds that
\begin{equation*}
\dmaxeps{\rho^{\otimes n}}{\sigma^{\otimes n}}{\eps} = n\relent{\rho}{\sigma} + \sqrt{nV(\rho\|\sigma)} \Phi^{-1}(\eps) + O(\log n) ,
\end{equation*}
and 
\begin{equation*}
\dmineps{\rho^{\otimes n}}{\sigma^{\otimes n}}{\eps} = n\relent{\rho}{\sigma} + \sqrt{nV(\rho\|\sigma)} \Phi^{-1}(\eps) + O(\log n) .
\end{equation*}
\end{fact}

Following fact can be viewed as a triangle inequality for smooth max-relative entropy.

\begin{fact}
\label{triangledmax}
For $\rho_A \in \mathcal{D}(A),\sigma_A,\tau_A\in\mathcal{P}(A)$, it holds that
$$\dmaxeps{\rho_A}{\tau_A}{\eps} \leq \dmaxeps{\rho_A}{\sigma_A}{\eps} + \dmax{\sigma_A}{\tau_A}.$$
\end{fact}
\begin{proof}
Let $k\defeq \dmax{\sigma_A}{\tau_A}$, which implies that $\sigma_A \preceq 2^k\tau_A$. Let $\rho'_A\in\ball{\eps}{\rho_A}$ be the state achieving the infimum in $R\defeq \dmaxeps{\rho_A}{\sigma_A}{\eps}$. Then $\rho'_A \preceq 2^R\sigma_A \preceq 2^{R+k}\tau_A$. This implies that $\dmax{\rho'_A}{\tau_A}\leq R+k$, which concludes the fact using the inequality $\dmaxeps{\rho_A}{\tau_A}{\eps} \leq \dmax{\rho'_A}{\tau_A}$.

\end{proof}

Following is an important fact that relates the information spectrum relative entropy to the max-relative entropy.

\begin{fact}[Lemma 12 and Proposition 13,\cite{TomHay13}]
\label{dmaxepsdseps}
Let $\eps, \delta\in (0,1)$ such that $\eps^2+\delta <1$. For quantum state $\rho_A\in\mathcal{D}(A)$ and  $\sigma \in \mathcal{P}(A)$, it holds that 
\begin{eqnarray*}
&&\dseps{\rho_A}{\sigma_A}{1-\eps^2-\delta} - 2\log\frac{1}{\delta}-2 \leq \dmaxeps{\rho_A}{\sigma_A}{\eps} \leq\\ && \dseps{\rho_A}{\sigma_A}{1-\eps^2+\delta} + \log v(\sigma) + 2\log\frac{1}{\eps} + \log\frac{1}{\delta},
\end{eqnarray*}
where $v(\sigma_A)$ is the number of distinct eigenvalues of $\sigma_A$. It also holds that 
$$ \dsepsalt{\rho_A}{\sigma_A}{\eps^2+\delta} - 2\log\frac{1}{\delta} -2 \leq \dmaxeps{\rho_A}{\sigma_A}{\eps}.$$
\end{fact}

\begin{proof}
The first part is essentially that given in \cite{TomHay13} (Proposition 13 and Lemma 12). For the second part, we note that the proof in \cite{TomHay13} (Proposition 12, Equation $23$) directly proceeds for this case as well: setting $R\defeq  \dmaxeps{\rho_A}{\sigma_A}{\eps}$, it is shown that for any $\delta'>0$, it holds that 
$$\Tr(\rho_A\{\rho_A- 2^{R+\delta'}\sigma_A\}_{-}) \geq (\sqrt{1-\eps^2}-2^{-\delta'/2})^2.$$ Setting $\delta' = \log\frac{1}{\delta}$, the inequality follows. 
\end{proof}

Following Fact immediately follows from the above Fact.
\begin{fact}
\label{smoothdseps}
Let $\eps, \eps_1\in (0,1)$. For quantum states $\rho_A,\rho'_A\in\mathcal{D}(A)$ and  $\sigma_A,\sigma'_A \in \mathcal{P}(A)$, the following properties hold. 
\begin{enumerate}
\item If $\rho'_A \in \ball{\eps}{\rho_A}$, then 
\begin{eqnarray*}
\dseps{\rho'_A}{\sigma_A}{1-\eps^2-3\eps_1^2} &\leq& \dseps{\rho_A}{\sigma_A}{1-\eps_1^2} + \log v(\sigma) \\ &&+ 8\log\frac{1}{\eps_1}
\end{eqnarray*}
 and 
\begin{eqnarray*}
\dsepsalt{\rho'_A}{\sigma_A}{\eps^2+3\eps_1^2} &\leq& \dseps{\rho_A}{\sigma_A}{1-\eps_1^2} + \log v(\sigma)\\ && + 8\log\frac{1}{\eps_1}.
\end{eqnarray*}
\item 
\begin{eqnarray*}
\dsepsalt{\rho_A}{\sigma'_A}{4\eps_1^2} &\leq& \dseps{\rho_A}{\sigma_A}{1-\eps_1^2} + \dmax{\sigma_A}{\sigma'_A}\\ &&+\log v(\sigma) + 4\log\frac{1}{\eps_1}.
\end{eqnarray*}

\end{enumerate}
\end{fact}
\begin{proof}
The items are proved as follows.
\begin{enumerate}
\item  Consider the following series of inequalities, which follow by application of Fact \ref{dmaxepsdseps} and the observation that 
$\dmaxeps{\rho_A}{\sigma_A}{\delta} \geq \dmaxeps{\rho'_A}{\sigma_A}{\delta+\eps}$ for any $\delta>0$.

\begin{eqnarray*}
&&\dseps{\rho_A}{\sigma_A}{1-\eps_1^2} \geq \\ &&\dmaxeps{\rho_A}{\sigma_A}{\sqrt{2\eps_1}} - \log v(\sigma) - 4\log\frac{1}{\eps_1}\\ &&\geq \dmaxeps{\rho'_A}{\sigma_A}{\sqrt{2}\eps_1+\eps} - \log v(\sigma) - 4\log\frac{1}{\eps_1}\\ &&\geq  \dseps{\rho'_A}{\sigma_A}{1-3\eps_1^2-\eps^2} - \log v(\sigma) - 8\log\frac{1}{\eps_1}.
\end{eqnarray*}

Second expression of the item follows similarly.

\item Let $k\defeq \dmax{\sigma_A}{\sigma'_A}$. Consider the following series of inequalities, which follow by the application of Fact \ref{dmaxepsdseps} and Fact \ref{triangledmax}.

\begin{eqnarray*}
&&\dseps{\rho_A}{\sigma_A}{1-\eps_1^2} \\ && \geq\dmaxeps{\rho_A}{\sigma_A}{\sqrt{2\eps_1}} - \log v(\sigma) - 4\log\frac{1}{\eps_1}\\ &&\geq \dmaxeps{\rho_A}{\sigma'_A}{\sqrt{2}\eps_1} - k- \log v(\sigma) - 4\log\frac{1}{\eps_1}\\ &&\geq \dsepsalt{\rho_A}{\sigma'_A}{4\eps_1^2} - k-\log v(\sigma) - 4\log\frac{1}{\eps_1}.
\end{eqnarray*}
\end{enumerate}
\end{proof}

Following fact relates the two definitions of the information spectrum relative entropy used in our proofs.

\begin{fact}
\label{dsepsrelate}
Fix an $\eps\in (0,1)$. For quantum state $\rho_A\in \mathcal{D}(A)$ and operator $\sigma_A\in \mathcal{P}(A)$, we have that 
$$\dseps{\rho_A}{\sigma_A}{\eps} \leq \dsepsalt{\rho_A}{\sigma_A}{1-\eps}.$$
\end{fact}
\begin{proof}
Let $R$ achieve the infimum in the definition of $\dsepsalt{\rho_A}{\sigma_A}{1-\eps}$. Thus, we have that $\Tr(\rho_A\{\rho_A-2^R\sigma_A\}_-) \geq \eps$. Using the relation $\{\rho_A-2^R\sigma_A\}_-  ~~+~~ \{\rho_A-2^R\sigma_A\}_+ = \id_A$, we obtain $\Tr(\rho_A\{\rho_A-2^R\sigma_A\}_+) < 1-\eps$. From the fact that $\Tr(\rho_A\{\rho_A-2^R\sigma_A\}_+)$ is monotonically decreasing in $R$ (as shown in \cite{NagaokaH07}, Equation 17), we conclude the proof.
\end{proof}

Following two facts are about some special properties of information spectrum relative entropy.
\begin{fact}
\label{projecteddseps}
Let $\eps \in (0,1)$. Let $\rho_A,\sigma_A \in \mathcal{D}(A)$ be quantum states such that $\rho_A \in \mathrm{supp}(\Pi_A)$ for some projector $\Pi_A$ that commutes with $\sigma_A$. Then it holds that,
$$\dseps{\rho_A}{\sigma_A}{\eps} = \dseps{\rho_A}{\Pi_A\sigma_A\Pi_A}{\eps}.$$
\end{fact}
\begin{proof}
For any $k>0$, it holds that $$\rho_A-2^k\sigma_A = \Pi_A\rho_A\Pi_A - 2^k \Pi_A\sigma_A\Pi_A - 2^k (\id_A-\Pi_A)\sigma_A(\id_A-\Pi_A).$$ Thus, 
\begin{eqnarray*}
\{\rho_A-2^k\sigma_A\}_+ &=& \{\Pi_A\rho_A\Pi_A - 2^k \Pi_A\sigma_A\Pi_A\}_+ \\ &=& \{\rho_A - 2^k \Pi_A\sigma_A\Pi_A\}_+. 
\end{eqnarray*}
This completes the proof by using the definition of information spectrum relative entropy.
\end{proof}

\begin{fact}
\label{purestatedseps}
Let $\eps \in (0,1)$, Let $\rho_A\in \mathcal{D}(A)$ be a pure quantum state and $\mu_A$ be the maximally mixed state on register $A$. Then
$$\dseps{\rho_A}{\mu_A}{\eps} = \dmax{\rho_A}{\mu_A}.$$
\end{fact}
\begin{proof}
By definition, we have $\dseps{\rho_A}{\mu_A}{\eps} = \sup{R:\Tr(\rho_A\{\rho_A-2^R\mu_A\}_+)\geq 1-\eps}$. Now the projector $\{\rho_A-2^R\mu_A\}_+$ contains the support of $\rho_A$ if $R < \log\mathrm{dim}(A)$ and otherwise is a null projector. The constraint $\Tr(\rho_A\{\rho_A-2^R\mu_A\}_+) \geq 1-\eps > 0$ requires that $\{\rho_A-2^R\mu_A\}_+$ be a non-null projector. This implies that $\dseps{\rho_A}{\mu_A}{\eps} = \log\mathrm{dim}(A)$, which is equal to the value of $\dmax{\rho_A}{\mu_A}$. This completes the proof.
\end{proof}

Following fact relates a projection on one system of a bipartite pure state to a projection on the other system.

\begin{fact}
\label{transpose}
Let $\rho_{AB} \in \mathcal{D}(AB)$ be a pure quantum state. Let $\Pi_A$ be a projector on the support of $\rho_A$ that commutes with $\rho_A$. Then there exists a projector $\Pi_B$ acting on register $B$ (which we refer to as a dual to the projector $\Pi_A$) such that $\Pi_A\rho_{AB}\Pi_A = \Pi_B\rho_{AB}\Pi_B$, $\Pi_B$ commutes with $\rho_B$ and belongs to the support of $\rho_B$.
\end{fact}
\begin{proof}
Let $d$ be the dimension of the subspace corresponding to $\Pi_A$. Since $\Pi_A$ and $\rho_A$ commute, there exists a basis $\{\ket{e_i}_A\}_{i=1}^{|A|}$ on register $A$ and a basis $\{\ket{f_i}_B\}_{i=1}^{|B|}$ on register $B$ such that 
$$\Pi_A = \sum_{i=1}^d \ketbra{e_i}_A, \quad \ket{\rho}_{AB} = \sum_{i}\lambda_i\ket{e_i}_A\ket{f_i}_B.$$ Define $\Pi_B \defeq \sum_{i=1}^d \ketbra{f_i}_B$. It can be verified that $\Pi_B$ satisfies the properties mentioned in the statement.  
\end{proof}

We shall also use the well known Chernoff bounds.
\begin{fact}[Chernoff bounds]
\label{fact:chernoff}
Let $\eps \in (0,1)$. Let $X_1 , \ldots, X_n$ be independent random variables, with each $X_i \in [0,1]$ always. Let $X \defeq X_1 + \cdots + X_n$ and  $\mu \defeq \frac{\mathbb{E} X}{n} = \frac{\mathbb{E} X_1 + \cdots + \mathbb{E} X_n}{n}$. Then
\begin{align*}
\mathrm{Pr}(X \geq n(\mu+\varepsilon)) &\leq \exp\left(-n\frac{\varepsilon^2}{3\mu}\right) \\
\mathrm{Pr}(X \leq n(\mu-\varepsilon)) &\leq \exp\left(-n\frac{\varepsilon^2}{2\mu}\right)  .
\end{align*}
\end{fact}

\subsection*{Statement of the main theorem}

Now we proceed to the main result of this section, which shows that given a pure state $\rho$, one can find a pure state $\psi$ close to $\rho$ that satisfies several constraints on the max-relative entropy. 

\begin{theorem}
\label{thm:asymptotic}
Let $\delta \in (0, \frac{1}{6000})$. Let $\rho_{RMN}\in \mathcal{D}(RMN)$ be a pure quantum state. Fix an integer $n$ such that: \begin{eqnarray*}
n &>&10^5\cdot\log\frac{2}{\delta}\cdot\max\bigg\{ \frac{S(\rho_M)\cdot\log(1/\lambda_{min}(\rho_M))}{\delta^2},\\ && \frac{S(\rho_N)\cdot\log(1/\lambda_{min}(\rho_N))}{\delta^2},\\ && \frac{S(\rho_R)\cdot\log(1/\lambda_{min}(\rho_R))}{\delta^2}\bigg\}.
\end{eqnarray*}

Then there exists a pure quantum state $\psi_{R^nM^nN^n}$ such that
\begin{enumerate}
\item $\Pur(\psi_{R^nM^nN^n},\rho^{\otimes n}_{RMN})\leq 60\sqrt{\delta}$.
\item $\psi_{R^n} \preceq (1+2000\delta)\rho_{R}^{\otimes n}$.
\item $ \dmax{\psi_{R^nN^n}}{\rho^{\otimes n}_{R}\otimes \rho^{\otimes n}_{N}} \leq \dmaxeps{\rho^{\otimes n}_{RN}}{\rho^{\otimes n}_{R}\otimes \rho^{\otimes n}_{N}}{\sqrt{\delta}} + 10\log\frac{1}{\delta} + 12n\delta+ O(\log n)$.
\item $ \dmax{\psi_{R^nM^n}}{\rho^{\otimes n}_{R}\otimes \rho^{\otimes n}_{M}} \leq \dmaxeps{\rho^{\otimes n}_{RM}}{\rho^{\otimes n}_{R}\otimes \rho^{\otimes n}_{M}}{\sqrt{\delta}} + 10\log\frac{1}{\delta} + 12n\delta + O(\log n)$.
\item $\dmax{\psi_{R^nM^nN^n}}{\rho^{\otimes n}_{R}\otimes\rho^{\otimes n}_{M}\otimes \rho^{\otimes n}_{N}}\leq \dmaxeps{\rho^{\otimes n}_{RMN}}{\rho^{\otimes n}_{R}\otimes\rho^{\otimes n}_{M}\otimes\rho^{\otimes n}_{N}}{\sqrt{\delta}} + 10\log\frac{1}{\delta} +  12n\delta +  O(\log n).$
\end{enumerate}
\end{theorem}

\subsection*{A warm-up lemma}

We first prove a simpler version of Theorem \ref{thm:asymptotic}, wherein we assume that each of the marginals of the quantum state $\rho_{RMN}$ are maximally mixed. More formally, we show the following lemma (note that the statement below is in fact in one-shot). 
\begin{lemma}
\label{lem:unifoneshot}
Let $\delta\in (0,\frac{1}{5})$. Let $\rho_{RMN} \in \mathcal{D}(RMN)$ be a pure quantum state such that $\rho_R= \frac{\id_R}{|R|}, \rho_M = \frac{\id_M}{|M|}, \rho_N = \frac{\id_N}{|N|}$. Then there exists a pure quantum state $\rho''_{RMN} \in \ball{5\delta}{\rho_{RMN}}$ such that
\begin{enumerate}
\item $\rho''_R \preceq \frac{\rho_R}{1-10\delta^2}$.
\item $\dmax{\rho''_{RN}}{\rho_R\otimes \rho_N} \leq \dmaxeps{\rho_{RN}}{\rho_R\otimes \rho_N}{\delta} + 6\log\frac{1}{\delta}$.
\item $\dmax{\rho''_{RM}}{\rho_R\otimes \rho_M} \leq \dmaxeps{\rho_{RM}}{\rho_R\otimes \rho_M}{\delta} + 6\log\frac{1}{\delta}$.
\item $\dmax{\rho''_{RMN}}{\rho_R\otimes \rho_M \otimes \rho_N} \leq \dmaxeps{\rho_{RMN}}{\rho_R\otimes \rho_M \otimes \rho_N}{\delta} + 21\log\frac{1}{\delta}$.
\end{enumerate}
\end{lemma} 
\begin{proof}
Below, we use the fact that maximally mixed quantum states have exactly one eigenvalue. From Fact \ref{dmaxepsdseps} (which relates the information spectrum relative entropy to the max-relative entropy) and Fact \ref{dsepsrelate} (which relates the two definitions of information spectrum relative entropy), we conclude the following relations.
\begin{eqnarray}
\label{eq:oneshotrhoprop}
&&\dseps{\rho_{RMN}}{\rho_R\otimes \rho_M\otimes \rho_N}{1-2\delta^2}\nonumber\\ &&\leq \dsepsalt{\rho_{RMN}}{\rho_R\otimes \rho_M\otimes \rho_N}{2\delta^2} \nonumber\\ &&\leq \dmaxeps{\rho_{RMN}}{\rho_R\otimes \rho_M\otimes \rho_N}{\delta} + 5\log\frac{1}{\delta}, \nonumber\\
&&\dseps{\rho_{RM}}{\rho_R\otimes \rho_M}{1-2\delta^2} \leq \dsepsalt{\rho_{RM}}{\rho_R\otimes \rho_M}{2\delta^2}\nonumber \\ &&\leq \dmaxeps{\rho_{RM}}{\rho_R\otimes \rho_M}{\delta} + 5\log\frac{1}{\delta},\nonumber\\
&&\dseps{\rho_{RN}}{\rho_R\otimes \rho_N}{1-2\delta^2} \leq \dsepsalt{\rho_{RN}}{\rho_R\otimes \rho_N}{2\delta^2}\nonumber\\ && \leq \dmaxeps{\rho_{RN}}{\rho_R\otimes \rho_N}{\delta} + 5\log\frac{1}{\delta}.
\end{eqnarray}

Let $k$ be the minimum achieved in $\dsepsalt{\rho_{RM}}{\rho_R\otimes \rho_M}{2\delta^2}$. Let $\Pi\defeq \{\rho_{RM} - 2^k\rho_R\otimes\rho_M\}_{-}$ and define the state 
$$\rho'_{RMN}\defeq \frac{\Pi\rho_{RMN}\Pi}{\Tr(\Pi\rho_{RMN})}.$$
It holds that $\Tr(\Pi\rho_{RM})\geq 1-2\delta^2$. We prove the following properties of $\rho'_{RMN}$. First item shows that $\rho'$ is close to $\rho$. Second and third items show that $\rho'$ now satisfies the desired max-relative entropy constraints on systems $R$ and $R,M$. Fourth item shows that $\rho'$ still retains the information spectrum relative entropy properties of $\rho$.  
\begin{claim}
\label{oneshotrhoprime}
It holds that
\begin{enumerate}
\item $\Pur(\rho'_{RMN}, \rho_{RMN}) \leq \sqrt{2}\delta$.
\item $\rho'_R\preceq \frac{\rho_R}{1-2\delta^2}$.
\item $\dmax{\rho'_{RM}}{\rho_R\otimes\rho_M} \leq  \dsepsalt{\rho_{RM}}{\rho_R\otimes \rho_M}{2\delta^2} + \log\frac{1}{1-2\delta^2}$.
\item $\dsepsalt{\rho'_{RN}}{\rho_R\otimes \rho_N}{8\delta^2} \leq \dmaxeps{\rho_{RN}}{\rho_R\otimes \rho_N}{\delta} + 13\log\frac{1}{\delta}.$
\item $\dsepsalt{\rho'_{RMN}}{\rho_R\otimes \rho_M\otimes \rho_N}{8\delta^2} \leq \dmaxeps{\rho_{RMN}}{\rho_R\otimes \rho_M\otimes \rho_N}{\delta} + 13\log\frac{1}{\delta}.$
\end{enumerate}
\end{claim}
\begin{proof}We prove each item as follows.
\begin{enumerate}
\item From the Gentle measurement lemma \ref{gentlelemma}, we have $\F^2(\rho'_{RMN}, \rho_{RMN}) \geq \Tr(\Pi\rho_{RMN}) \geq 1-2\delta^2$. Thus, $\Pur(\rho'_{RMN}, \rho_{RMN}) \leq \sqrt{2}\delta$. 
\item Since $\rho_R\otimes \rho_M$ commutes with $\rho_{RM}$, $\Pi$ commutes with $\rho_{RM}$ as well. Thus, 
$$\rho'_{RM}= \frac{\Pi\rho_{RM}\Pi}{\Tr(\Pi\rho_{RMN})} \preceq \frac{\rho_{RM}}{1-2\delta^2}.$$
Thus, we conclude the statement by tracing out register $M$.
\item By definition of $\Pi$, we have 
$$\rho'_{RM}= \frac{\Pi\rho_{RM}\Pi}{\Tr(\Pi\rho_{RMN})} \preceq \frac{2^k\rho_R\otimes \rho_M}{1-2\delta^2}.$$
This concludes the statement from the definition of $k$.
\item From Equation \ref{eq:oneshotrhoprop} and Fact \ref{smoothdseps} (which relates the information spectrum relative entropies of two close-by quantum states), we have 
\begin{eqnarray*}
&&\dsepsalt{\rho'_{RN}}{\rho_R\otimes \rho_N}{8\delta^2}\\ && \leq \dseps{\rho_{RN}}{\rho_R\otimes \rho_N}{1-2\delta^2} + 8\log\frac{1}{\delta^2}\\ &&\leq \dmaxeps{\rho_{RN}}{\rho_R\otimes \rho_N}{\delta} + 13\log\frac{1}{\delta}. 
\end{eqnarray*}
\item Similar arguments as above imply that
\begin{eqnarray*}
&& \dsepsalt{\rho'_{RMN}}{\rho_R\otimes \rho_M\otimes \rho_N}{8\delta^2} \\ &&\leq \dmaxeps{\rho_{RMN}}{\rho_R\otimes \rho_M\otimes \rho_N}{\delta} + 13\log\frac{1}{\delta}.
\end{eqnarray*}
\end{enumerate}
\end{proof}

Let $k'$ be the minimum achieved in $\dsepsalt{\rho'_{RN}}{\rho_R\otimes \rho_N}{8\delta^2}$. Let $\Pi'\defeq \{\rho'_{RN} - 2^{k'}\rho_R\otimes \rho_N\}_{-}$ and define the state 
$$\rho''_{RMN}\defeq \frac{\Pi'\rho'_{RMN}\Pi'}{\Tr(\Pi'\rho'_{RMN})}.$$ It holds that $\Tr(\Pi'\rho'_{RN}) \geq 1-8\delta^2$. We prove the following properties for $\rho''_{RMN}$. First property says that $\rho''$ is close to $\rho$. Second and third properties say that the max-relative entropy constraints on the registers $R$ and $R,N$ hold for $\rho''$. Fourth property says that the max-relative entropy constraint on $\rho'_{RM}$ continues to hold for $\rho''$ as well, even when the projector $\Pi'$ acted on a different subsystem. This uses Fact \ref{transpose} in its argument. Fifth property shows that the max-entropy constraint holds on all the registers $RMN$. 
\begin{claim}It holds that
\begin{enumerate}
\item $\Pur(\rho''_{RMN}, \rho_{RMN}) \leq 5\delta$.
\item $\rho''_R \preceq \frac{\rho_R}{1-10\delta^2}$.
\item $\dmax{\rho''_{RN}}{\rho_R\otimes \rho_N} \leq \dmaxeps{\rho_{RN}}{\rho_R\otimes \rho_N}{\delta} + 6\log\frac{1}{\delta}$.
\item $\dmax{\rho''_{RM}}{\rho_R\otimes \rho_M} \leq \dmaxeps{\rho_{RM}}{\rho_R\otimes \rho_M}{\delta} + 6\log\frac{1}{\delta}$.
\item $\dmax{\rho''_{RMN}}{\rho_R\otimes \rho_M \otimes \rho_N} \leq \dmaxeps{\rho_{RMN}}{\rho_R\otimes \rho_M \otimes \rho_N}{\delta} + 21\log\frac{1}{\delta}$.
\end{enumerate}
\end{claim}
\begin{proof}
We prove each item as follows.
\begin{enumerate}
\item This follows from the application of the Gentle measurement lemma \ref{gentlelemma}.
\item This follows similarly along the lines of Item $2$ in Claim \ref{oneshotrhoprime}.
\item This follows along the lines similar to Item $3$, Claim \ref{oneshotrhoprime}.
\item We note that the projector $\Pi'$ commutes with $\rho'_{RN}$ and belongs to its support. Since $\rho'_{RMN}$ is a pure state, there exists a dual projector $\tilde{\Pi}$ acting on register $M$ that belongs to the support of $\rho'_M$ and commutes with $\rho'_M$ (Fact \ref{transpose}). Further, $\tilde{\Pi}$ satisfies $\Tr(\tilde{\Pi}\rho'_M) = \Tr(\Pi''\rho'_{RN})$ and $\rho'' =  \frac{\tilde{\Pi}\rho'_{RMN}\tilde{\Pi}}{\Tr(\tilde{\Pi}\rho'_{RMN})}$. Thus, we find 

\begin{eqnarray*}
\rho''_{RM} &\preceq& \frac{\tilde{\Pi}\rho'_{RM}\tilde{\Pi}}{1-8\delta^2} \preceq \frac{2^k\tilde{\Pi}\rho_R\otimes \rho_M\tilde{\Pi}}{1-10\delta^2} \\ &=& \frac{2^k\rho_R\otimes \tilde{\Pi}\rho_M\tilde{\Pi}}{1-10\delta^2} \preceq \frac{2^k\rho_R\otimes \rho_M}{1-10\delta^2}.
\end{eqnarray*}

Above, the second inequality follows from Item $3$, Claim \ref{oneshotrhoprime}. This concludes the item, from the definition of $k$ and Equation \ref{eq:oneshotrhoprop}.
\item Along the lines similar to Item $5$, Claim \ref{oneshotrhoprime}, we have that 
\begin{eqnarray*}
&&\dsepsalt{\rho''_{RMN}}{\rho_R\otimes \rho_M\otimes \rho_N}{25\delta^2}\\ && \leq \dmaxeps{\rho_{RMN}}{\rho_R\otimes \rho_M\otimes \rho_N}{\delta} + 21\log\frac{1}{\delta}.
\end{eqnarray*}
But $\rho''_{RMN}$ is a pure quantum state. Thus from the choice of $\delta$, which ensures that $25\delta^2 <1$ and Fact \ref{purestatedseps} (which shows that the smooth max-relative entropy and the information spectrum relative entropy coincide for pure states), we obtain
\begin{eqnarray*}
&&\dmax{\rho''_{RMN}}{\rho_R\otimes \rho_M \otimes \rho_N} \\ &&\leq \dmaxeps{\rho_{RMN}}{\rho_R\otimes \rho_M \otimes \rho_N}{\delta} + 21\log\frac{1}{\delta}.
\end{eqnarray*}
\end{enumerate}
\end{proof}

This completes the proof of the lemma. 
\end{proof}

\subsection*{Proof of the main theorem}

Now we proceed to the proof of Theorem \ref{thm:asymptotic}. Its proof roughly follows the proof of Lemma \ref{lem:unifoneshot} above. Additional care is required in the arguments due to the fact that the marginals of $\rho_{RMN}^{\otimes n}$ are not exactly uniform.

\begin{proof}[Proof of Theorem \ref{thm:asymptotic}]
Our proof is divided into three main steps. 

\vspace{2mm}

\noindent {\bf Typical projection onto the subsystems $R$, $M$, $N$:} For brevity, we set $\rho_{R^nM^nN^n} \defeq \rho_{RMN}^{\otimes n}$. Let $\Pi_{R^n}$ be the projector onto the eigenvectors of $\rho_{R^n}$ with eigenvalues in the range $[2^{-n(S(\rho_R)+\delta) }, 2^{-n(S(\rho_R)-\delta)}]$. Similarly, define  $\Pi_{M^n}, \Pi_{N^n}$. Let $\mu_{R^n},\mu_{M^n},\mu_{N^n}$ be the uniform distributions in the support of $\Pi_{R^n},\Pi_{M^n}, \Pi_{N^n}$ respectively. Using Chernoff bounds (Fact \ref{fact:chernoff}),  we have that $$\Tr(\Pi_{R^n}\rho_{R^n})\geq 1-2\cdot \exp(-\frac{\delta^2\cdot n}{ S(\rho_R)\cdot\log(1/\lambda_{min}(\rho_R))}) \geq 1-\delta$$ for the choice of $n$. Similarly, $\Tr(\Pi_{M^n}\rho_{M^n})\geq 1-\delta$ and $\Tr(\Pi_{N^n}\rho_{N^n})\geq 1-\delta$.

Thus, the dimensions of the projectors $\Pi_{R^n},\Pi_{M^n}, \Pi_{N^n}$ are in the range $[(1-\delta)2^{n(S(\rho_R)-\delta)}, 2^{n(S(\rho_R)+\delta)}]$, $[(1-\delta)2^{n(S(\rho_M)-\delta)}, 2^{n(S(\rho_M)+\delta)}]$ and $[(1-\delta) 2^{n(S(\rho_N)-\delta)}, 2^{n(S(\rho_N)+\delta)}]$, respectively. Following relations are now easy to observe.
\begin{eqnarray}
\label{uniforms}
&& (1-\delta)2^{-2n\delta}\Pi_{R^n}\rho_{R^n}\Pi_{R^n} \preceq \mu_{R^n}\nonumber\\ && \preceq (1+\delta)2^{2n\delta}\Pi_{R^n}\rho_{R^n}\Pi_{R^n} \preceq (1+\delta)2^{2n\delta}\rho_{R^n}, \nonumber \\
&& (1-\delta)2^{-2n\delta}\Pi_{M^n}\rho_{M^n}\Pi_{M^n} \preceq \mu_{M^n}\nonumber\\ &&\preceq (1+\delta)2^{2n\delta}\Pi_{M^n}\rho_{M^n}\Pi_{M^n} \preceq (1+\delta)2^{2n\delta}\rho_{M^n}, \nonumber \\
&& (1-\delta)2^{-2n\delta}\Pi_{N^n}\rho_{N^n}\Pi_{N^n} \preceq \mu_{N^n} \nonumber\\ &&\preceq (1+\delta)2^{2n\delta}\Pi_{N^n}\rho_{N^n}\Pi_{N^n} \preceq (1+\delta)2^{2n\delta}\rho_{N^n}.
\end{eqnarray}
Now, define the state 
\begin{eqnarray*}
&&\rho'_{R^nM^nN^n} \defeq\\ && \frac{(\Pi_{R^n}\otimes \Pi_{M^n}\otimes \Pi_{N^n})\rho_{R^nM^nN^n}(\Pi_{R^n}\otimes \Pi_{M^n}\otimes \Pi_{N^n})}{\Tr(\rho_{R^nM^nN^n}(\Pi_{R^n}\otimes\Pi_{M^n}\otimes \Pi_{N^n}))}.
\end{eqnarray*}
 We will establish the following claims about $\rho'_{R^nM^nN^n}$.
\begin{claim}
\label{rhoprimeclaims}
It holds that 
\begin{enumerate}
\item $\F^2(\rho'_{R^nM^nN^n},  \rho_{R^nM^nN^n})\geq 1-64\delta$.
\item $\Tr(\rho_{R^nM^nN^n}(\Pi_{R^n}\otimes \Pi_{M^n}\otimes \Pi_{N^n})) \geq 1-64\delta$.
\item $\rho'_{R^nM^nN^n} \in \mathrm{supp}(\Pi_{R^n}\otimes \Pi_{M^n}\otimes \Pi_{N^n})$.
\item $\rho'_{R^n} \preceq \frac{1}{1-64\delta}\rho_{R^n}, \rho'_{M^n} \preceq \frac{1}{1-64\delta}\rho_{M^n},$\\ $\rho'_{N^n} \preceq \frac{1}{1-64\delta}\rho_{N^n}$.
\end{enumerate}
\end{claim}
\begin{proof}
We prove each item in a sequence below.
\begin{enumerate}
\item This is a straightforward application of Gentle measurement lemma (Fact \ref{gentlelemma}) and Fact \ref{closestatesmeasurement}.
\item This follows along similar lines as argued above.
\item This follows since $(\Pi_{R^n}\otimes \Pi_{M^n}\otimes \Pi_{N^n})\rho_{R^nM^nN^n}(\Pi_{R^n}\otimes \Pi_{M^n}\otimes \Pi_{N^n}) \preceq \Pi_{R^n}\otimes \Pi_{M^n}\otimes \Pi_{N^n}$.
\item We proceed as follows for $\rho'_{R^n}$.  
\begin{eqnarray*}
\rho'_{R^n} &=& \frac{1}{\Tr(\rho_{R^nM^nN^n}(\Pi_{R^n}\otimes \Pi_{M^n}\otimes \Pi_{N^n}))}\\ &&\Tr_{M^nN^n}((\Pi_{R^n}\otimes \Pi_{M^n}\otimes \Pi_{N^n})\rho_{R^nM^nN^n}\\ &&((\Pi_{R^n}\otimes \Pi_{M^n}\otimes \Pi_{N^n}))\\ &=& \frac{1}{\Tr(\rho_{R^nM^nN^n}(\Pi_{R^n}\otimes \Pi_{M^n}\otimes \Pi_{N^n}))}\\&&\Tr_{M^nN^n}((\Pi_{R^n}\otimes \Pi_{M^n}\otimes \Pi_{N^n})\rho_{R^nM^nN^n}\\ &&((\Pi_{R^n}\otimes \id_{M^n}\otimes \id_{N^n}))\\ &\preceq& \frac{1}{\Tr(\rho_{R^nM^nN^n}(\Pi_{R^n}\otimes \Pi_{M^n}\otimes \Pi_{N^n}))}\\ &&\Tr_{M^nN^n}((\Pi_{R^n}\otimes \id_{M^n}\otimes \id_{N^n})\rho_{R^nM^nN^n}\\ &&((\Pi_{R^n}\otimes \id_{M^n}\otimes \id_{N^n})) \\ &=& \frac{1}{\Tr(\rho_{R^nM^nN^n}(\Pi_{R^n}\otimes \Pi_{M^n}\otimes \Pi_{N^n}))}\Pi_{R^n}\rho_{R^n}\Pi_{R^n}  \\ &&\preceq \frac{1}{1-64\delta} \rho_{R^n}.
\end{eqnarray*}
Last inequality is due to Item $2$ above and the fact that $\Pi_{R^n}$ is a projector onto certain eigenspace of $\rho_{R^n}$. Same argument holds for $\rho_{M^n}$ and $\rho_{N^n}$.
\end{enumerate}
\end{proof}
\noindent{\bf Switching to the information spectrum relative entropy:} Using above claim, we now proceed to the second step of our proof. As a corollary from the Claim (Item $1$), along with Fact \ref{dmaxepsdseps}, we conclude 
\begin{eqnarray}
\label{switchofdmax}
&&\dseps{\rho'_{R^nM^nN^n}}{\rho_{R^n}\otimes\rho_{M^n}\otimes \rho_{N^n}}{1-90\delta}-2\log\frac{1}{\delta}\nonumber\\ &&\leq \dmaxeps{\rho'_{R^nM^nN^n}}{\rho_{R^n}\otimes \rho_{M^n}\otimes\rho_{N^n}}{9\sqrt{\delta}} \nonumber\\ &&\leq \dmaxeps{\rho_{R^nM^nN^n}}{\rho_{R^n}\otimes\rho_{M^n}\otimes \rho_{N^n}}{\sqrt{\delta}} .
\end{eqnarray}
Fact \ref{projecteddseps} implies that 
\begin{eqnarray*}
&&\dseps{\rho'_{R^nM^nN^n}}{\rho_{R^n}\otimes\rho_{M^n}\otimes \rho_{N^n}}{1-90\delta}=\\ &&  \mathrm{D}_s^{1-90\delta}\bigg(\rho'_{R^nM^nN^n} \| \Pi_{R^n}\rho_{R^n}\Pi_{R^n}\\&&\otimes\Pi_{M^n}\rho_{M^n}\Pi_{M^n}\otimes\Pi_{N^n}\rho_{N^n}\Pi_{N^n}\bigg).
\end{eqnarray*}

Let $v \defeq v(\Pi_{R^n}\rho_{R^n}\Pi_{R^n}\otimes\Pi_{M^n}\rho_{M^n}\Pi_{M^n}\otimes\Pi_{N^n}\rho_{N^n}\Pi_{N^n})$, which is the number of distinct eigenvalues of $\Pi_{R^n}\rho_{R^n}\Pi_{R^n}\otimes\Pi_{M^n}\rho_{M^n}\Pi_{M^n}\otimes\Pi_{N^n}\rho_{N^n}\Pi_{N^n}$. We apply Fact \ref{smoothdseps} along with Equation \ref{uniforms} to conclude that
\begin{eqnarray*}
 &&\mathrm{D}_s^{1-90\delta}\bigg(\rho'_{R^nM^nN^n}\|\Pi_{R^n}\rho_{R^n}\Pi_{R^n}\otimes\\ &&\Pi_{M^n}\rho_{M^n}\Pi_{M^n}\otimes\Pi_{N^n}\rho_{N^n}\Pi_{N^n}\bigg) \\ && \geq \dsepsalt{\rho'_{R^nM^nN^n}}{\mu_{R^n}\otimes\mu_{M^n}\otimes \mu_{N^n}}{400\delta}\\ && - \log\frac{2^{6n\delta}}{(1-\delta)^3} - \log v - 5\log\frac{1}{\delta}.
\end{eqnarray*}
Combining this with Equation \ref{switchofdmax}, we conclude that
\begin{eqnarray}
\label{uniformandactualRMN}
&&\dsepsalt{\rho'_{R^nM^nN^n}}{\mu_{R^n}\otimes\mu_{M^n}\otimes \mu_{N^n}}{400\delta}\nonumber\\ &&\leq \dmaxeps{\rho_{R^nM^nN^n}}{\rho_{R^n}\otimes \rho_{M^n}\otimes\rho_{N^n}}{\sqrt{\delta}} \nonumber \\ &&+ 8\log\frac{1}{\delta} + 6n\delta+ \log v.
\end{eqnarray}

In the same way, we can argue that 
\begin{eqnarray}
\label{uniformandactualRM}
&&\dsepsalt{\rho'_{R^nM^n}}{\mu_{R^n}\otimes\mu_{M^n}}{400\delta}\nonumber\\ &&\leq \dmaxeps{\rho_{R^nM^n}}{\rho_{R^n}\otimes \rho_{M^n}}{\sqrt{\delta}} \nonumber\\ &&+ 8\log\frac{1}{\delta} + 6n\delta +  \log v.
\end{eqnarray}
and 
\begin{eqnarray}
\label{uniformandactualRN}
&&\dsepsalt{\rho'_{R^nN^n}}{\mu_{R^n}\otimes\mu_{N^n}}{400\delta}\nonumber \\ &&\leq \dmaxeps{\rho_{R^nN^n}}{\rho_{R^n}\otimes \rho_{N^n}}{\sqrt{\delta}} \nonumber\\ && + 8\log\frac{1}{\delta} + 6n\delta +  \log v.
\end{eqnarray}
{\bf Removing large eigenvalues from a subsystem:} Let $k$ be the minimum achieved in $\dsepsalt{\rho'_{R^nM^n}}{\mu_{R^n}\otimes\mu_{M^n}}{400\delta}$. For brevity, set $\Pi' \defeq \{\rho'_{R^nM^n}-2^{k}\mu_{R^n}\otimes\mu_{M^n}\}_{-}$ and define the state 
$$\rho''_{R^nM^nN^n}\defeq \frac{\Pi'\rho'_{R^nM^nN^n}\Pi'}{\Tr(\Pi'\rho'_{R^nM^nN^n})}.$$ It holds that $\Tr(\Pi'\rho'_{A^nB^n}) \geq 1 - 400\delta$. We prove the following properties for $\rho''_{R^nM^nN^n}$.
\begin{claim}
\label{rhodoubleprimeclaims} It holds that
\begin{enumerate}
\item $\Pur(\rho''_{R^nM^nN^n},\rho_{R^nM^nN^n})\leq 30\sqrt{\delta}$.
\item $\rho''_{R^n} \preceq (1+1000\delta)\rho_{R^n}, \rho''_{M^n} \preceq (1+1000\delta)\rho_{M^n}, \rho''_{N^n} \preceq (1+1000\delta)\rho_{N^n}$. Furthermore, $\rho''_{R^n} \in \mathrm{supp}(\Pi_{R^n})$, $\rho''_{M^n} \in \mathrm{supp}(\Pi_{M^n})$ and $\rho''_{N^n} \in \mathrm{supp}(\Pi_{N^n})$.
\item $ \dmax{\rho''_{R^nM^n}}{\rho_{R^n}\otimes \rho_{M^n}} \leq \dmaxeps{\rho_{R^nM^n}}{\rho_{R^n}\otimes \rho_{M^n}}{\sqrt{\delta}} + 9\log\frac{1}{\delta} + 12n\delta + \log v$.
\item $\dsepsalt{\rho''_{R^nM^nN^n}}{\mu_{R^n}\otimes\mu_{M^n}\otimes \mu_{N^n}}{1300\delta}\leq \dmaxeps{\rho_{R^nM^nN^n}}{\rho_{R^n}\otimes \rho_{M^n}\otimes\rho_{N^n}}{\sqrt{\delta}} + 8\log\frac{1}{\delta} + 6n\delta +\log v.$\newline
$\dsepsalt{\rho''_{R^nN^n}}{\mu_{R^n}\otimes\mu_{N^n}}{1300\delta}\leq \dmaxeps{\rho_{R^nN^n}}{\rho_{R^n}\otimes \rho_{N^n}}{\sqrt{\delta}} + 8\log\frac{1}{\delta} + 6n\delta +\log v.$
\end{enumerate}
\end{claim}

\begin{proof}
We prove the items in the respective sequence.
\begin{enumerate}

\item From Gentle measurement lemma \ref{gentlelemma}, we have that $\F^2(\rho''_{R^nM^nN^n},\rho'_{R^nM^nN^n})\geq \Tr(\Pi'\rho'_{R^nM^nN^n}) \geq 1-400\delta$. Using Claim \ref{rhoprimeclaims} (Item 1) and triangle inequality for purified distance (Fact \ref{fact:trianglepurified}), we obtain that $\Pur(\rho''_{R^nM^nN^n},\rho_{R^nM^nN^n})\leq 30\sqrt{\delta}$. 

\item  Since $\mu_{R^n}\otimes \mu_{M^n}$ is uniform in the support of $\rho'_{R^nM^n}$, $\rho'_{R^nM^n}$ commutes with $\mu_{R^n}\otimes \mu_{M^n}$. This immediately implies that $\Pi'$ commutes with $\rho'_{R^nM^n}$. Thus, we conclude that $$\rho''_{R^nM^n} = \frac{\Pi'\rho'_{R^nM^n}\Pi'}{\Tr(\Pi'\rho'_{R^nM^n})} \preceq \frac{\rho'_{R^nM^n}}{1-400\delta},$$ where the inequality follows from the relation $\Tr(\Pi'\rho'_{R^nM^n})\geq 1-400\delta$.

Invoking Claim \ref{rhoprimeclaims} (Item 4), we obtain $$\rho''_{R^n} \preceq \frac{\rho'_{R^n}}{1-400\delta} \preceq \frac{\rho_{R^n}}{(1-400\delta)(1-10\delta)} \preceq \frac{\rho_{R^n}}{1-410\delta}.$$ Similarly, we obtain $\rho''_{M^n} \preceq \frac{\rho_{M^n}}{1-410\delta}$. The inequality $\rho''_{N^n} \preceq \frac{\rho'_{N^n}}{1-400\delta}$ follows from the fact that $\Pi'$ does not act on register $N^n$. First part of the item now follows since $\frac{1}{1-410\delta} < 1+1000\delta$ for the choice of $\delta$. 

For the second part, we use the fact that $\rho'_{R^n} \in \mathrm{supp}(\Pi_{R^n})$ and the relation $\rho''_{R^n} \preceq \frac{\rho'_{R^n}}{1-400\delta}$ established above. Same argument holds for $\rho''_{M^n}, \rho''_{N^n}$.

\item By definition of $\Pi'$, we have that $$\Pi'\rho'_{R^nM^n}\Pi' \preceq 2^{k}\Pi'\mu_{R^n}\otimes \mu_{M^n}\Pi' \preceq 2^{k}\mu_{R^n}\otimes \mu_{M^n},$$ where last inequality holds since $\mu_{R^n}\otimes \mu_{M^n}$ is uniform and $\Pi'$ is in its support. Thus, $$\rho''_{R^nM^n} = \frac{\Pi'\rho'_{R^nM^n}\Pi'}{\Tr(\Pi'\rho'_{R^nM^n})}\preceq \frac{2^{k}}{1-400\delta}\cdot\mu_{R^n}\otimes\mu_{M^n}.$$ 

 From Equation \ref{uniforms}, this further implies that
$$\rho''_{R^nM^n} \preceq \frac{(1+\delta)^2\cdot 2^{2n\delta}\cdot 2^{k}}{1-400\delta}\cdot\rho_{R^n}\otimes\rho_{M^n}.$$ This proves the item after using Equation \ref{uniformandactualRM} to upper bound $k$.

\item Applying Fact \ref{dsepsrelate}, we conclude from Equation \ref{uniformandactualRMN} that 
\begin{eqnarray*}
&&\dseps{\rho'_{R^nM^nN^n}}{\mu_{R^n}\otimes\mu_{M^n}\otimes \mu_{N^n}}{1-400\delta}\\ &&\leq \dmaxeps{\rho_{R^nM^nN^n}}{\rho_{R^n}\otimes \rho_{M^n}\otimes\rho_{N^n}}{\sqrt{\delta}}\\ && + 8\log\frac{1}{\delta} + 6n\delta + \log v.
\end{eqnarray*} 
Now we use Fact \ref{smoothdseps} along with the Item 1 above, which says that $\Pur(\rho''_{R^nM^nN^n},\rho_{R^nM^nN^n})\leq 30\sqrt{\delta}$, to conclude that
\begin{eqnarray*}
&&\dsepsalt{\rho''_{R^nM^nN^n}}{\mu_{R^n}\otimes\mu_{M^n}\otimes \mu_{N^n}}{1300\delta}\\ &&\leq \dmaxeps{\rho_{R^nM^nN^n}}{\rho_{R^n}\otimes \rho_{M^n}\otimes\rho_{N^n}}{\sqrt{\delta}}\\&& + 8\log\frac{1}{\delta} + 6n\delta +\log v.
\end{eqnarray*}
Second expression in this item follows similarly using Equation \ref{uniformandactualRN}.
\end{enumerate}
\end{proof}
\noindent {\bf Removing large eigenvalues from another subsystem:} Let $k'$ be the minimum achieved in \newline
$\dsepsalt{\rho''_{R^nN^n}}{\mu_{R^n}\otimes\mu_{N^n}}{1300\delta}.$ For brevity, set $\Pi'' \defeq \{\rho''_{R^nN^n}-2^{k'}\mu_{R^n}\otimes\mu_{N^n}\}_{-}$ and define the state 
$$\rho'''_{R^nM^nN^n}\defeq \frac{\Pi''\rho''_{R^nM^nN^n}\Pi''}{\Tr(\Pi''\rho''_{R^nM^nN^n})}.$$ It holds that $\Tr(\Pi''\rho''_{A^nB^n}) \geq 1 - 1300\delta$. We prove the following properties for $\rho'''_{R^nM^nN^n}$.

\begin{claim}
\label{rhotripleprimeclaims} It holds that
\begin{enumerate}
\item $\Pur(\rho'''_{R^nM^nN^n},\rho_{R^nM^nN^n})\leq 60\sqrt{\delta}$.
\item $\rho'''_{R^n} \preceq (1+2000\delta)\rho_{R^n}$.
\item $ \dmax{\rho'''_{R^nN^n}}{\rho_{R^n}\otimes \rho_{N^n}} \leq \dmaxeps{\rho_{R^nN^n}}{\rho_{R^n}\otimes \rho_{N^n}}{\sqrt{\delta}} + 10\log\frac{1}{\delta} + 12n\delta + \log v$.
\item $ \dmax{\rho'''_{R^nM^n}}{\rho_{R^n}\otimes \rho_{M^n}} \leq \dmaxeps{\rho_{R^nM^n}}{\rho_{R^n}\otimes \rho_{M^n}}{\sqrt{\delta}} + 10\log\frac{1}{\delta} + 12n\delta +\log v$.
\item $\dmax{\rho'''_{R^nM^nN^n}}{\rho_{R^n}\otimes\rho_{M^n}\otimes \rho_{N^n}}\leq \dmaxeps{\rho_{R^nM^nN^n}}{\rho_{R^n}\otimes\rho_{M^n}\otimes\rho_{N^n}}{\sqrt{\delta}} + 10\log\frac{1}{\delta} + 12n\delta +\log v.$
\end{enumerate}
\end{claim}
\begin{proof}
We prove each item as follows.
\begin{enumerate}
\item This follows similarly along the lines of Item 1 in Claim \ref{rhodoubleprimeclaims}.
\item This follows similarly along the lines of Item 2 in Claim \ref{rhodoubleprimeclaims}.
\item This follows similarly along the lines of Item 3 in Claim \ref{rhodoubleprimeclaims} and also uses Item 4 of Claim \ref{rhodoubleprimeclaims}.
\item We note that the projector $\Pi''$ commutes with $\rho''_{R^nN^n}$ and belongs to its support. Since $\rho''_{R^nM^nN^n}$ is a pure state, there exists a dual projector $\tilde{\Pi}$ acting on register $M^n$ that belongs to the support of $\rho''_{M^n}$ and commutes with $\rho''_{M^n}$ such that $\Tr(\tilde{\Pi}\rho''_{R^nM^nN^n}) = \Tr(\Pi''\rho''_{R^nM^nN^n})$ and $\rho'''_{R^nM^nN^n} = \frac{\tilde{\Pi}\rho''_{R^nM^nN^n}\tilde{\Pi}}{\Tr(\tilde{\Pi}\rho''_{R^nM^nN^n})}$. 

Thus, we find 
\begin{eqnarray*}
&&\rho'''_{R^nM^n} \preceq \frac{\tilde{\Pi}\rho''_{R^nM^n}\tilde{\Pi}}{1-1300\delta} \preceq \frac{2^k \tilde{\Pi}\mu_{R^n}\otimes \mu_{M^n}\tilde{\Pi}}{1-1700\delta} \\ && = \frac{2^k \mu_{R^n}\otimes \tilde{\Pi}\mu_{M^n}\tilde{\Pi}}{1-1700\delta} \preceq \frac{2^k \mu_{R^n}\otimes \mu_{M^n}}{1-1700\delta}.
\end{eqnarray*}

Above, the second inequality is immediate from the definition of $\rho''_{R^nM^n}$ (and also appears in the proof of Item 3, Claim \ref{rhodoubleprimeclaims}). Last inequality follows from the observation that $\tilde{\Pi}$ is in the support of $\rho''_{M^n}$ and hence in the support of $\Pi_{M^n}$, as implied by Claim \ref{rhodoubleprimeclaims} (Item 2). This ensures that $\tilde{\Pi}$ is in the support of $\mu_{M^n}$. The item concludes by using Equation \ref{uniforms}.
\item From Claim \ref{rhodoubleprimeclaims} (Item 4), and Fact \ref{smoothdseps}, we conclude that 
\begin{eqnarray*}
&&\dseps{\rho'''_{R^nM^nN^n}}{\mu_{R^n}\otimes\mu_{M^n}\otimes \mu_{N^n}}{1-5000\delta}\\ &&\leq \dmaxeps{\rho_{R^nM^nN^n}}{\rho_{R^n}\otimes \rho_{M^n}\otimes\rho_{N^n}}{\sqrt{\delta}}\\&& + 10\log\frac{1}{\delta} + 6n\delta+ \log v.
\end{eqnarray*}

But observe that $\rho'''_{R^nM^nN^n}$ is a pure state. Thus, from the choice of $\delta$ which ensures that $5000\delta < 1$ and Fact \ref{purestatedseps}, we have that 
\begin{eqnarray*}
&&\dmax{\rho'''_{R^nM^nN^n}}{\mu_{R^n}\otimes\mu_{M^n}\otimes \mu_{N^n}}\\ &&\leq \dmaxeps{\rho_{R^nM^nN^n}}{\rho_{R^n}\otimes \rho_{M^n}\otimes\rho_{N^n}}{\sqrt{\delta}} \\ &&+ 10\log\frac{1}{\delta} + 6n\delta+\log v.
\end{eqnarray*} 
The item now follows from Equation \ref{uniforms}.
\end{enumerate}
\end{proof}

Now the value of $v$, which is the number of distinct eigenvalues of $\Pi_{R^n}\rho_{R^n}\Pi_{R^n}\otimes \Pi_{M^n}\rho_{M^n}\Pi_{M^n}\otimes \Pi_{N^n}\rho_{N^n}\Pi_{N^n}$, is upper bounded by the number of distinct eigenvalues of $\rho_{R^n}\otimes \rho_{M^n}\otimes \rho_{N^n}$. This is at most $n^{2|R|+2|M|+2|N|}$. This proves the theorem, combining with the Claim \ref{rhotripleprimeclaims} and setting $\psi_{R^nM^nN^n} = \rho'''_{R^nM^nN^n}$.
\end{proof}

\end{document}